\documentclass[aps,showpacs,twocolumn,longbibliography,nofootinbib,secnumarabic]{revtex4-1}
\usepackage{graphicx}
\usepackage{amssymb}
\usepackage{float}
\usepackage[T1]{fontenc}
\usepackage{amsmath}
\usepackage{xcolor}
\begin{document}
\title
{\bf  A binary mixture of Bose-Einstein-condensates in a double-well potential: Berry phase and two-mode entanglement}
\author{Mehmet G\"{u}nay} \email{gunaymehmt@gmail.com}
\affiliation{Department of Nanoscience and Nanotechnology, Faculty of Arts and Science, Mehmet Akif Ersoy University, 15030 Burdur, Turkey}
\break

\begin{abstract}
A binary mixtures of Bose-Einstein condensate~(BEC) structures exhibit an incredible richness in terms of holding different kinds of phases. Depending on the ratio of the inter- and intra-atomic interactions, the transition from mixed to separated phase, which is also known as the miscibility-immiscibility transition, has been reported in different setups.
Here, we describe such type of quantum phase transition~(QPT) in an effective Hamiltonian approach, by applying Holstein-Primakoff transformation in the limit of large number of particles. We demonstrate that a non-trivial geometric phase near the critical coupling is present, which confirms the connection between Berry phase and QPT. We also show that, by using the spin form of Hillery $\&$ Zubairy criterion, a two-mode entanglement accompanies this transition in the limit of large, but not infinite number of particles.

\end{abstract}
\maketitle
 \section{Introduction}
In recent years, there has been growing interest in studying two-component quantum fluids. Phase mixing and separation
of the two components due to relative strength of inter-species and intra-species interactions open gate to investigate variety of research topics including the dynamical phase transitions~\cite{PhysRevA.81.043613,PhysRevA.94.013602,mertes2007nonequilibrium}, the production of dipolar molecules~\cite{PhysRevLett.113.255301}, two-mode entanglement~\cite{PhysRevA.71.013601,PhysRevLett.106.120405}, and the macroscopic quantum self trapping~\cite{levy2007ac,zibold2010classical}. 
The phase separation phenomenon was first observed in $ ^{3}{\rm He}$ - $^{4}{\rm He}$ mixtures~\cite{PhysRev.103.262.2}. Later, it was reported in different BEC structures~\cite{myatt1997production,PhysRevLett.81.1539,PhysRevA.82.033609,PhysRevLett.101.040402,simoni2003magnetic,roy2017two,PhysRevA.84.011603,lercher2011production}. The ability of the resonant control of two-body interactions via Feshbach resonances makes these structures attractive for practical applications. 

The observation of \textit{controllable} phase separation of a binary BEC was reported by using Feshbach resonance in the hyperfine levels~\cite{myatt1997production,PhysRevLett.81.1539,PhysRevA.82.033609} and in the isotopes~\cite{PhysRevLett.101.040402} of the rubidium atoms. Additionally, the similar results were also obtained by using different kinds of atoms~\cite{simoni2003magnetic,roy2017two,PhysRevA.84.011603,lercher2011production}.
Theoretical investigations of these structures have shown that the relative interaction strength and number of particles play a crucial role in characterizing the density profiles. A two-species Bose-Hubbard~(BH) model, in the limit of a weakly interacting gas, is widely used due to performing a fully analytic derivation~\cite{PhysRevE.95.062142}. The characterization of the self-trapping~\cite{PhysRevLett.79.4950}, entanglement~\cite{lingua2016delocalization,lingua2018residual}, the dynamical phase transition~\cite{PhysRevA.74.063614}, etc., have been done within the BH model.

In this work, we theoretically study the collective behaviors of two-species BEC trapped in a double-well potential. While the dynamical properties of these structures have been extensively studied through the mean field approach~\cite{PhysRevA.81.043613,tian2016dynamical} and the well-described Bose-Hubbard model~\cite{PhysRevA.71.013601,PhysRevA.84.023629} that analyze the low-excitations with the use of Holstein-Primakoff~(HP) transformation~\cite{holstein1940field,PhysRevA.11.981}, the analytical descriptions were resided mainly in the mixed-phase solutions.
The motivation of the present paper is to obtain simple analytical solutions to describe the system in each of its mixed and separated phases, which can serve to investigate the quantum properties of such mixtures. In that respect, we study the Berry phase and bipartite entanglement through the miscibility-immiscibility transition.

In many-body systems, the quantum phase transition~(QPT) can be observed when crossing between ground and excited states takes place. It is known that such level crossings generate singularities in the Hilbert space, and is therefore natural to expect the reflections of such behaviors in the wavefunction. The Berry phase can capture these points, and its connection with QPT have been studied in different models~\cite{carollo2005geometric,PhysRevE.74.031123,plastina2006scaling}. Here, we obtain a non-trivial geometric phase by encircling the critical point and observe that with the increasing number of particles, the transition in the value of the Berry phase becomes sharper around the critical point, and in the thermodynamic limit, N$ \to \infty $, there appears a step-like behavior. Having the ability to adiabatically control the interatomic interaction strength between two species~\cite{PhysRevA.82.033609,PhysRevLett.101.040402}, makes this result valuable for the Berry phase-related applications, and it also provides a tool to detect criticality in the presence of QPT.

Besides fundamental interest, the model studied here also offers the possibility to test bipartite entanglement from macroscopic observables. In large systems, it was shown that entanglement can be inferred from collective spin measurements~\cite{giovannetti2003characterizing,sorensen2001Nature}, and experimental observations using this method have been reported between two spatially separated  atomic ensembles~\cite{EsslingerNature2010Dicke,riedel2010atom,matsukevich2006entanglement,simon2007single,fadel2018spatial,kunkel2018spatially,lange2018entanglement} and between the spins of atoms in optical lattices~\cite{PhysRevLett.115.035302,islam2015measuring}. Entanglement characterization of the similar models are, in general, done by using von Neumann entropy \cite{vidal2007entanglement}. Here we analyze such phenomena through the miscibility-immiscibility transition, by adopting the \textit{spin form} of two-mode entanglement witness~\cite{PhysRevA.86.023626}, originally introduced for two-mode entanglement in ~\cite{HZPRL2006}. We observe that the criterion witnesses the entanglement onset in the separated phase. Unlike the Berry phase, entanglement decays faster with the increasing number of particles and/or interspecies interaction strength. We find that this is due to the vanishing effective coupling term, which is responsible for two-mode squeezing.

The manuscript is organized as follows. In Sec.~\ref{sec-model}, we introduce the model of a two-species BECs trapped in a double-well potential and derive the effective Hamiltonians and associated ground state wavefunctions of the mixed and separated phases. The appearance of the nontrivial geometric phase around the critical coupling is observed in Sec.~\ref{sec-BP}. In Sec.~\ref{sec-Ent}, we discuss the formation of bipartite entanglement by anticipating spin form of Hillery $\&$ Zubairy criterion. A  summary appears in Sec.~\ref{sec-sum}.
\pagebreak

 \section{The model} \label{sec-model}
We consider a two species~($a$ and $b$) condensate mixtures trapped in a double-well potential with large number of particles~N$ _{a(b)} \gg$ 1 and in the low-excitation limit. By assuming the trap frequency~($\omega$) of each local potential is much larger than the interactions among the atoms~(N$_i g_{i}$), i.e. $\omega \gg$N$_i g_i$ with $i=a,b$, we construct the Hamiltonian with two-mode approximation~\cite{PhysRevA.55.4318,PhysRevA.73.013602,PhysRevA.71.013601}, which is given by~\cite{PhysRevA.81.043613}
 \begin{eqnarray}
\hat{\cal H} &=& \frac{ t_a}{2} (\hat{a}_L^\dagger  \hat{a}_R+\hat{a}_R^\dagger  \hat{a}_L)+ \frac{ t_b}{2} (\hat{b}_L^\dagger  \hat{b}_R+\hat{b}_R^\dagger  \hat{b}_L) \nonumber\\
&+& \frac{g_a}{2} [(\hat{a}_L^\dagger  \hat{a}_L )^2 +( \hat{a}_R^\dagger  \hat{a}_R  )^2  ] +\frac{ g_b  }{2} [(\hat{b}_L^\dagger  \hat{b}_L )^2 +( \hat{b}_R^\dagger  \hat{b}_R  )^2  ] \nonumber \\
&+& g_{ab} (\hat{a}_L^\dagger  \hat{a}_L \hat{b}_L^\dagger  \hat{b}_L+\hat{a}_R^\dagger  \hat{a}_R \hat{b}_R^\dagger  \hat{b}_R). \label{Eq:Ham_BEC}
 \end{eqnarray}
Here, $ \hat{a}_j^\dagger $ ($ \hat{a}_j $) and $ \hat{b}_j^\dagger $ ($ \hat{b}_j $) are the creation (annihilation) operators of the species $a$ and $b$ respectively that residing in the $j$th well, $j=$ $L$, $R$. The parameters $t_{a} $ and $t_{b} $ describe the coupling~(tunneling) between two wells, $ g_{a(b)} $ and $ g_{ab} $ stand for intraspecies and interspecies interaction strength respectively, which is explicitly given by: $ g_{\alpha \beta}= 2 \pi \hbar^2 A_{\alpha \beta}/m_{\alpha \beta} 
\int|\phi_\alpha|^2|\phi_\beta|^2 dr$~\cite{tian2016dynamical}. Here $A_{\alpha \beta}$ is the $s$-wave scattering length between atoms, $ m_{\alpha \beta} $ is the reduced mass, where we denote $ g_\alpha=g_{\alpha \alpha} $ and $\alpha,\beta=a,b$.

The analysis of the Hamiltonian in Eq.~(\ref{Eq:Ham_BEC}) can be  simplified by the introduction of the angular momentum operators for each species as~\cite{sakurai2014modern}:
 \begin{eqnarray}
\hat{J}_{\alpha x}&=& (\hat{\alpha}_L^\dagger  \hat{\alpha}_L-\hat{\alpha}_R^\dagger  \hat{\alpha}_R)/2, \nonumber \\
\hat{J}_{\alpha y}&=& (\hat{\alpha}_L^\dagger  \hat{\alpha}_R-\hat{\alpha}_R^\dagger  \hat{\alpha}_L)/2i, \nonumber \\
\hat{J}_{\alpha z}&=& (\hat{\alpha}_L^\dagger  \hat{\alpha}_R+\hat{\alpha}_R^\dagger  \hat{\alpha}_L)/2, 
\label{Eq:Angular}
 \end{eqnarray}
where $ \alpha=a,b $. These operators obey the usual angular momentum commutation relations: $ [\hat{J}_{\alpha}^{+},\hat{J}_{\alpha}^{-}]=2 \hat{J}_{\alpha z} $ and  $ [\hat{J}_{\alpha}^{\pm},\hat{J}_{\alpha z}]= \mp \hat{J}_{\alpha}^{\pm} $, where $\hat{J}_{\alpha}^{\pm} =\hat{J}_{\alpha x}\pm i \hat{J}_{\alpha y} $. Inserting these definitions into the Eq.~(\ref{Eq:Ham_BEC}), the Hamiltonian can be rewritten as~\cite{PhysRevA.71.013601}
 \begin{eqnarray}
\hat{\cal H} &=& \sum_{\alpha=a,b} \{t_\alpha \hat{J}_{\alpha z}+g_\alpha  \hat{J}^2_{\alpha x}\}+ 2 g_{ab} \hat{J}_{ax}\hat{J}_{b x}. 
\label{Eq:Ham_Angular}
 \end{eqnarray}
In the limit of a large number of particles, one can make use of the Holstein-Primakoff~(HP) representation of the angular momentum operators. In this representation, the operators defined in Eq.~(\ref{Eq:Angular}) can be written in terms of a bosonic mode in the following way~\cite{holstein1940field,PhysRevA.11.981} 
\begin{eqnarray}
\hat{J}_{\alpha}^{+}=\hat{c}_{\alpha}^{\dagger} \sqrt{{\rm N}_\alpha-\hat{c}_{\alpha}^{\dagger}\hat{c}_{\alpha}}, && \quad \hat{J}_{\alpha}^{-}=\sqrt{{\rm N}_\alpha-\hat{c}_{\alpha}^{\dagger}\hat{c}_{\alpha}} \hat{c}_{\alpha}, \nonumber \\
\label{Eq:Hols_Prim} \\
\centering \hat{J}_{\alpha z}&=&\hat{c}_{\alpha}^{\dagger}\hat{c}_{\alpha}-j_\alpha, \nonumber
\end{eqnarray} 
where $j_\alpha=  {\rm N}_\alpha/2$, $ \alpha=a,b $ and $\hat{c}_{\alpha}  $ is the standard bosonic operator, having commutator $ [\hat{c}_{\alpha},\hat{c}^\dagger_{\alpha^\prime}] = \delta_{\alpha,\alpha^\prime}$. Next, we apply HP transformation and show that the Hamiltonian of two-spin system in Eq.~(\ref{Eq:Ham_Angular}) can be written in terms of two coupled oscillators~\cite{ZhengPRA2011ensemble_ensemble,gunay2019entanglement}. To do this, we insert Eq.~(\ref{Eq:Hols_Prim}) into the Eq.~(\ref{Eq:Ham_Angular}) and obtain

\begin{widetext}
\begin{eqnarray}
\hat{\cal H} &=& \sum_{\alpha=a,b} \Bigg \{ t_{\alpha} (\hat{c}_{\alpha}^{\dagger}\hat{c}_{\alpha}-{\rm N}_\alpha/2)   
+ \frac{g_\alpha j_\alpha}{2}\bigg( \hat{c}_{\alpha}^{\dagger} \sqrt{1-\frac{\hat{c}_{\alpha}^{\dagger}\hat{c}_{\alpha}}{{\rm N}_\alpha}}+\sqrt{1-\frac{\hat{c}_{\alpha}^{\dagger}\hat{c}_{\alpha}}{{\rm N}_\alpha}} \hat{c}_{\alpha}\bigg)^2 \Bigg \} \nonumber \\
&+& g_{ab}\sqrt{j_a j_b} \bigg(\hat{c}_{a}^{\dagger} \sqrt{1-\frac{\hat{c}_{a}^{\dagger}\hat{c}_{a}}{{\rm N}_a}}+\sqrt{1-\frac{\hat{c}_{a}^{\dagger}\hat{c}_{a}}{{\rm N}_a}} \hat{c}_{a}\bigg) 
\times  \bigg(\hat{c}_{b}^{\dagger} \sqrt{1-\frac{\hat{c}_{b}^{\dagger}\hat{c}_{b}}{{\rm N}_b}}+\sqrt{1-\frac{\hat{c}_{b}^{\dagger}\hat{c}_{b}}{{\rm N}_b}} \hat{c}_{b}\bigg).\label{Bosonic_Ham_BEC} 
\end{eqnarray} 
\end{widetext}

In the thermodynamic limit, N$_\alpha \to \infty$, one can obtain the effective Hamiltonian as  
\begin{eqnarray}
\hat{\cal H}^{(1)} &=& \sum_{\alpha=a,b} \{ t_{\alpha} (\hat{c}_{\alpha}^{\dagger}\hat{c}_{\alpha}-{\rm N}_\alpha/2) + \frac{g_\alpha j_\alpha}{2}( \hat{c}_{\alpha}^{\dagger} + \hat{c}_{\alpha})^2 \} \nonumber \\
&+& g_{ab}\sqrt{j_a j_b} (\hat{c}_{a}^{\dagger}+ \hat{c}_{a})(\hat{c}_{b}^{\dagger} + \hat{c}_{b}),\label{Bosonic_Ham_BEC_1} 
\end{eqnarray} 
which is analogous to that of two coupled oscillators. By defining the position and the momentum operators:
 \begin{eqnarray}
\hat{x}_a&=&  \frac{1}{\sqrt{2}}  (\hat{c}_{a}^{\dagger}+\hat{c}_{a}), \qquad \hat{p}_a= i  \frac {1}{\sqrt{2}}  (\hat{c}_{a}^{\dagger}-\hat{c}_{a}),  \label{Eq-posx}\\
\hat{x}_b&=&  \frac{1}{\sqrt{2}}  (\hat{c}_{b}^{\dagger}+\hat{c}_{b}), \qquad \hat{p}_b= i  \frac {1}{\sqrt{2}}  (\hat{c}_{b}^{\dagger}-\hat{c}_{b}),  \label{Eq-posy}
\end{eqnarray}  
we rewrite the effective Hamiltonian as
 \begin{eqnarray}
\hat{\cal H}^{(1)} &=&\sum_{\alpha=a,b} \big \{ \frac{t_\alpha}{2}(\hat{p}^2_\alpha + \hat{x}^2_\alpha)+\tilde{g}_\alpha    \hat{x}^2_\alpha \big \} + 2 \tilde{g}_{ab} \hat{x}_a \hat{x}_b,
\end{eqnarray}
where $ \tilde{g}_\alpha   = {g}_\alpha j_\alpha $ and $ \tilde{g}_{ab}   = {g}_{ab} \sqrt{j_a j_b }$. It is then straightforward to solve the resulting the normal mode frequencies of the coupled oscillation, which is given by
\begin{eqnarray}
\epsilon^{(1)}_\pm=\sqrt{\omega_+\pm \sqrt{\omega^2_-+4\tilde{g}_{ab} t_a t_b}}
\end{eqnarray}
where $ \omega_\pm =\tilde{g}_a t_a \pm \tilde{g}_b t_b+ (t_a^2\pm t_b^2)/2$. Crucially, one can see that the normal mode frequencies can have complex values depending on the interaction strengths, and the critical interspecies coupling strength ${g}_{ab}^{*}$ can be given as
\begin{eqnarray}
{g}_{ab}^{*} \equiv \frac{1}{2}\sqrt{\bigg(\frac{t_a}{j_a} +2g_a \bigg)\bigg(\frac{t_b}{j_b}+2g_b\bigg)}. \label{Eq:crit}
\end{eqnarray}
When the interspecies coupling strength exceeds this value, the system becomes unstable. Therefore, the condition, ${g}_{ab}>{g}_{ab}^{*}$~(${g}_{ab}<{g}_{ab}^{*}$), for the immisciblitiy~(misciblity) of the two species is satisfied. In the thermodynamic limit, $ {j_\alpha}\to \infty $, it reduces to well-known criticality (e.g. ${g}_{ab}^{*}= \sqrt{g_a g_b} $) for the  phase separation of two component BECs~\cite{PhysRevLett.77.3276}. Depending on the ratio of the intraspecies ~(${g}_{a(b)}$) and interspecies (${g}_{ab}$) interaction strengths, different types of the phases has been extensively studied theoretically~\cite{PhysRevA.59.634,PhysRevA.94.013602,PhysRevA.87.013625}, and such phases also seen experimentally~\cite{PhysRevA.84.011603,PhysRevA.82.033609}.

  \begin{figure}
\vskip-0.1truecm
\begin{center}
\includegraphics[width=80mm,height=40mm]{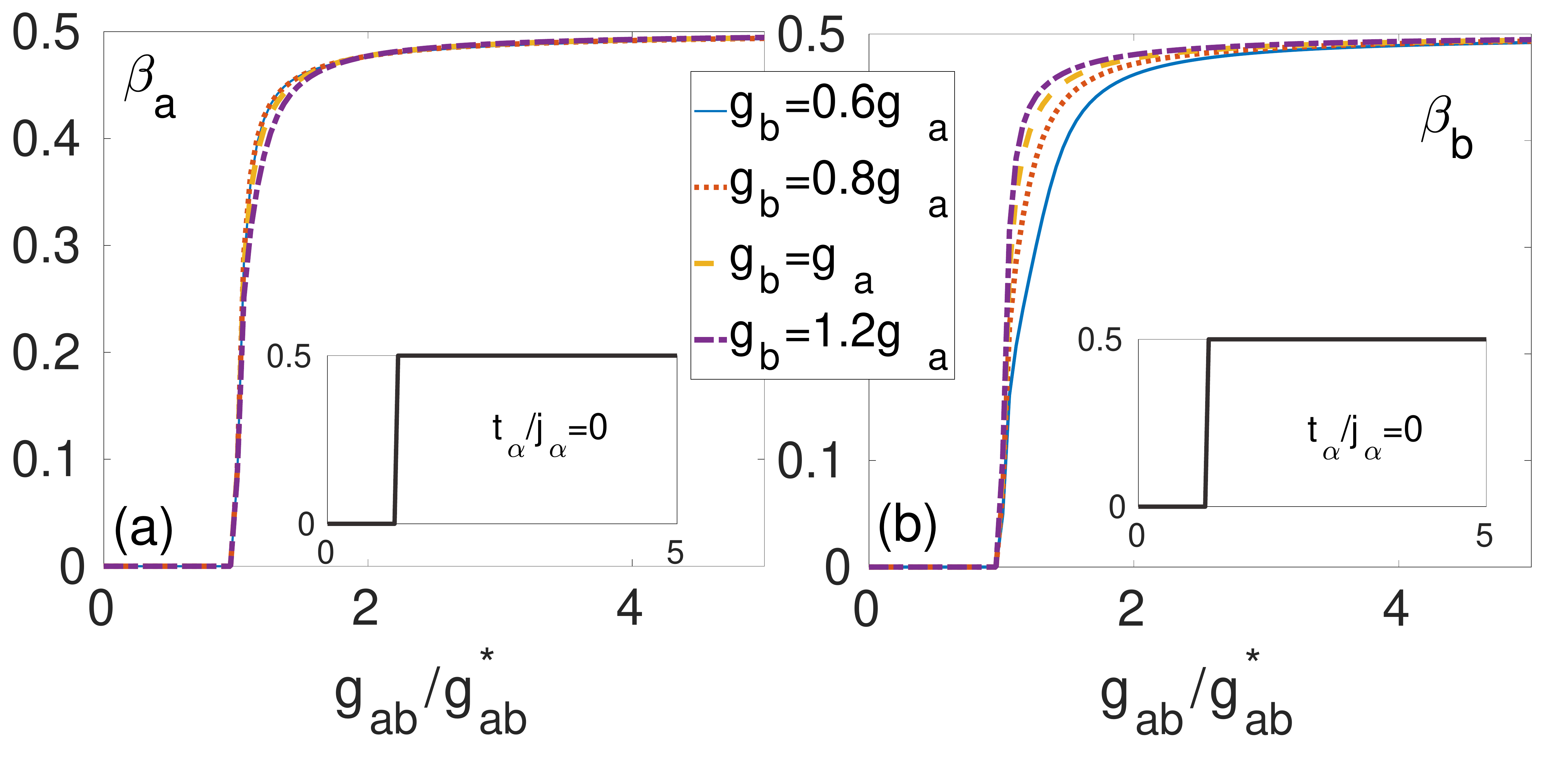}
\caption{The displacement amounts (a) $ \beta_a $ and (b) $ \beta_b $ as a function of coupling strength ${g}_{ab}$, for different values of ${g}_{b}$. Here we use equal tunneling amplitudes; $ {t}_{\alpha}/j_\alpha=0.1{g}_{a}$ and in the insets, we consider infinite number of particles case and take $ {t}_{\alpha}/{j}_{\alpha}=0$.}
\label{fig1}
\end{center}
\end{figure}

The main motivation in this work is to find a solution above a critical point, where phase transition occurs. Such type of phase transition is well-known in the Dicke-type models, in which one can describe the interaction of a single-mode quantized field with an ensemble of N two-level atoms. It was shown that above a critical coupling strength these systems can undergo quantum phase transition~\cite{wang1973phase,hepp1973superradiant,emary2003chaos}, from normal to \textit{superradiant phase}. Here we adopt this method~\cite{emary2003chaos,lambert2004entanglement} to characterize the phase transition between mixed and separated phases.

To find a solution in a region $ {g}_{ab}>{g}_{ab}^{*}$, we displace the bosonic operators as
 \begin{eqnarray}
\hat{c}_a^{\dagger}= \hat{d}_a^{\dagger} \pm\sqrt{{\rm N}_a \beta_a}, \qquad  \hat{c}_b^{\dagger}= \hat{d}_b^{\dagger} \mp\sqrt{{\rm N}_b \beta_b}. \label{Displacement}
\end{eqnarray}
In the following, we shall just consider the displacements as; $\hat{c}_a^{\dagger}= \hat{d}_a^{\dagger} +\sqrt{{\rm N}_a \beta_a} $ and $ \hat{c}_b^{\dagger}= \hat{d}_b^{\dagger} -\sqrt{{\rm N}_b \beta_b} $. If we insert these definitions into Eq.~(\ref{Bosonic_Ham_BEC}) and eliminate the first order term in the boson operators, one can find the amounts of displacement of each mode by solving 
 \begin{eqnarray}
 \frac{t_a}{j_a}+2 \biggl( g_a -{g}_{ab} \sqrt{\frac{\beta_b(1-\beta_b)}{\beta_a(1-\beta_a)}} \biggr)(1-2\beta_a)& =&0,  \label{a} \\ 
\frac{t_b}{j_b}+2 \biggl( g_b -{g}_{ab} \sqrt{\frac{\beta_a(1-\beta_a)}{\beta_b(1-\beta_b)}} \biggr)(1-2\beta_b) &=& 0.  \label{b}
\end{eqnarray}
The resulting effective Hamiltonian can be given by
\begin{eqnarray}
\hat{\cal H}^{(2)} &=& \sum_{\alpha=a,b} \{ \omega_\alpha \hat{d}_{\alpha}^{\dagger}\hat{d}_{\alpha}+\kappa_\alpha ( \hat{d}_{\alpha}^{\dagger} + \hat{d}_{\alpha})^2 \} \nonumber \\
&+& \lambda (\hat{d}_{a}^{\dagger}+ \hat{d}_{a})(\hat{d}_{b}^{\dagger} + \hat{d}_{b}),\label{Bosonic_Ham_BEC_2} 
\end{eqnarray} 
where we consider the boson operators up to the second order, and the parameters can be found as
 \begin{eqnarray}
\omega_\alpha&=& {t}_{\alpha}+2 {g}_{ab} j_{\overline{\alpha}} \sqrt{\beta_a \beta_b} \sqrt{\frac{1-\beta_{\overline{\alpha}}}{1-\beta_{\alpha}}}, \\
 \kappa_{\alpha}&=&\omega_\alpha-{t}_{\alpha} +{g}_{\alpha} j_\alpha \frac{6\beta_{\alpha}(\beta_\alpha-1)+1}{2(1-\beta_{\alpha})}, \\ 
\lambda &=& {g}_{ab} \sqrt{j_a j_b} \frac{(1-2 \beta_{a})(1-2 \beta_{b})}{\sqrt{(1-\beta_{a})(1-\beta_{b})}},  \label{Eq:eff_coup}
 \end{eqnarray}
here $\overline{a}=b  $ and $\overline{b}=a  $. Before proceeding, let us check these parameters in the limits of $ g_{ab}\leq g_{ab}^*$ and $ g_{ab} \gg g_{ab}^*$. For the case; $ g_{ab}\leq g_{ab}^*$ the displacement parameters have zero value, i.e. $ \beta_{\alpha}=0 $, and the Hamiltonian in Eq.~(\ref{Bosonic_Ham_BEC_2}) reduces to the one given for the mixed phase in Eq.~(\ref{Bosonic_Ham_BEC_1}). When $ g_{ab} \gg g_{ab}^*$, the displacement parameters take a single value, i.e. $ \beta_{\alpha}=0.5 $, as seen in FIG.~\ref{fig1}. In this limit, two oscillators become uncoupled,~$ \lambda=0 $ and by increasing interaction strength of the interspecies, it only contributes to the \textit{effective} strengths of the intraspecies,~$ \kappa_\alpha $, and tunneling,~$ \omega_\alpha $, coefficients, which makes the Hamiltonian $ \hat{\cal H}^{(2)} $ stable for all $ g_{ab}$ values. To see this, we find the new eigen-frequencies, by moving to a position-momentum representation defined by 
 \begin{eqnarray}
\hat{X}_a&=&  \frac{1}{\sqrt{2}}  (\hat{d}_{a}^{\dagger}+\hat{d}_{a}), \qquad \hat{P}_{a}= i  \frac {1}{\sqrt{2}}  (\hat{d}_{a}^{\dagger}-\hat{d}_{a}),  \label{Eq-posx}\\
\hat{X}_b&=&  \frac{1}{\sqrt{2}}  (\hat{d}_{b}^{\dagger}+\hat{d}_{b}), \qquad \hat{P}_b= i  \frac {1}{\sqrt{2}}  (\hat{d}_{b}^{\dagger}-\hat{d}_{b}),  \label{Eq-posy}
\end{eqnarray}  
where one can see the relation between the coordinates: $ \hat{X}_\alpha=\hat{x}_\alpha - \chi_\alpha \sqrt{2 {\rm N}_\alpha \beta_\alpha}$, $\chi_{a,b}=\pm 1$. By following the same steps, as it is done for $ \hat{\cal H}^{(1)} $ in Eq.~(\ref{Bosonic_Ham_BEC_1}), one can find the corresponding oscillator energies for $ \hat{\cal H}^{(2)} $ as 
 \begin{eqnarray}
  \epsilon_\pm^{(2)}=\frac{1}{2}\Bigl[\tilde{\omega}_a^2+\tilde{\omega}_b^2 \pm \sqrt{(\tilde{\omega}_a^2-\tilde{\omega}_b^2)^2+16{\lambda}^2 \omega_a \omega_b} \Bigr]^{1/2}
 \end{eqnarray} 
where $ \tilde{\omega}_\alpha^2= {\omega}_\alpha( \omega_\alpha + 4 {\kappa}_\alpha)$. The new excitation energy, $\epsilon_-^{(2)}  $, is real over the all parameter space and hence $ \hat{\cal H}^{(2)} $ describes the system in the phase separation region. 

Next, we give the ground state wavefunctions of the mixed and separated phases. Since the two effective Hamiltonians are obtained in the form of two coupled harmonic oscillators, their wavefunctions can be found as the product of harmonic oscillator eigenfunctions, which can be given by
 \begin{eqnarray}
\psi_{_{\rm GS}}^{(j)}(q_{ja},q_{jb})= \sqrt{\frac{m \Omega}{\pi}} G_1(q_{ja},q_{jb}) G_2(q_{ja},q_{jb}),
 \end{eqnarray}
where $ j=1,2 $ stands for the solutions of the mixed and separated phases respectively with $q_{_{1\alpha}}= x_\alpha $, $q_{_{2\alpha}}= X_\alpha  $ and $  G_1(q_{ja},q_{jb}) $ and $  G_2(q_{ja},q_{jb}) $ represent Gaussian functions defined by
 \begin{eqnarray}
G_1(q_{ja},q_{jb})&=& {e}^{\frac{m \Omega}{2}[\xi(q_{ja} { C}-q_{jb} { S})^2]} \\
G_2(q_{ja},q_{jb})&=& {e}^{\frac{m \Omega}{2}[\xi^{-1}(q_{ja} { S}+q_{jb} { C})^2]}
 \end{eqnarray}
with parameters
  \begin{eqnarray}
\xi &=& \frac{c_a+c_b+\sqrt{(c_a-c_b)^2+4 \lambda^2}}{2K},\\
  K&=&\sqrt{c_a c_b-\lambda/2}, \quad c_\alpha=(\omega_\alpha+4 \kappa_\alpha)\sqrt{\omega_{\overline{\alpha}}/\omega_\alpha},\\
 C&=& {\rm cos}(\phi), \quad S= {\rm sin}(\phi), \quad  {\rm tan}(2\phi)=\frac{2 \lambda}{c_a-c_b}, \\
 \Omega&=&\sqrt{K/m}, \quad {\rm and} \quad m=1/\sqrt{\omega_a \omega_b},
 \end{eqnarray}
where we denote $\overline{a}=b  $ and $\overline{b}=a  $. We define the parameters above for only  $ \psi_{_{\rm GS}}^{(2)} $, by inserting $ \beta_\alpha=0 $ into the these parameters one can obtain desired solution for  $ \psi_{_{\rm GS}}^{(1)} $.

In the following sections, since having derived the effective Hamiltonians and associated ground states that describe mixed and separated phases, we investigate the quantum features of this kind of phase transition in terms of geometric phase and bipartite entanglement. 

\section{Geometric phase}\label{sec-BP} 

\begin{figure}
\begin{center}
\includegraphics[width=86mm,height=45mm]{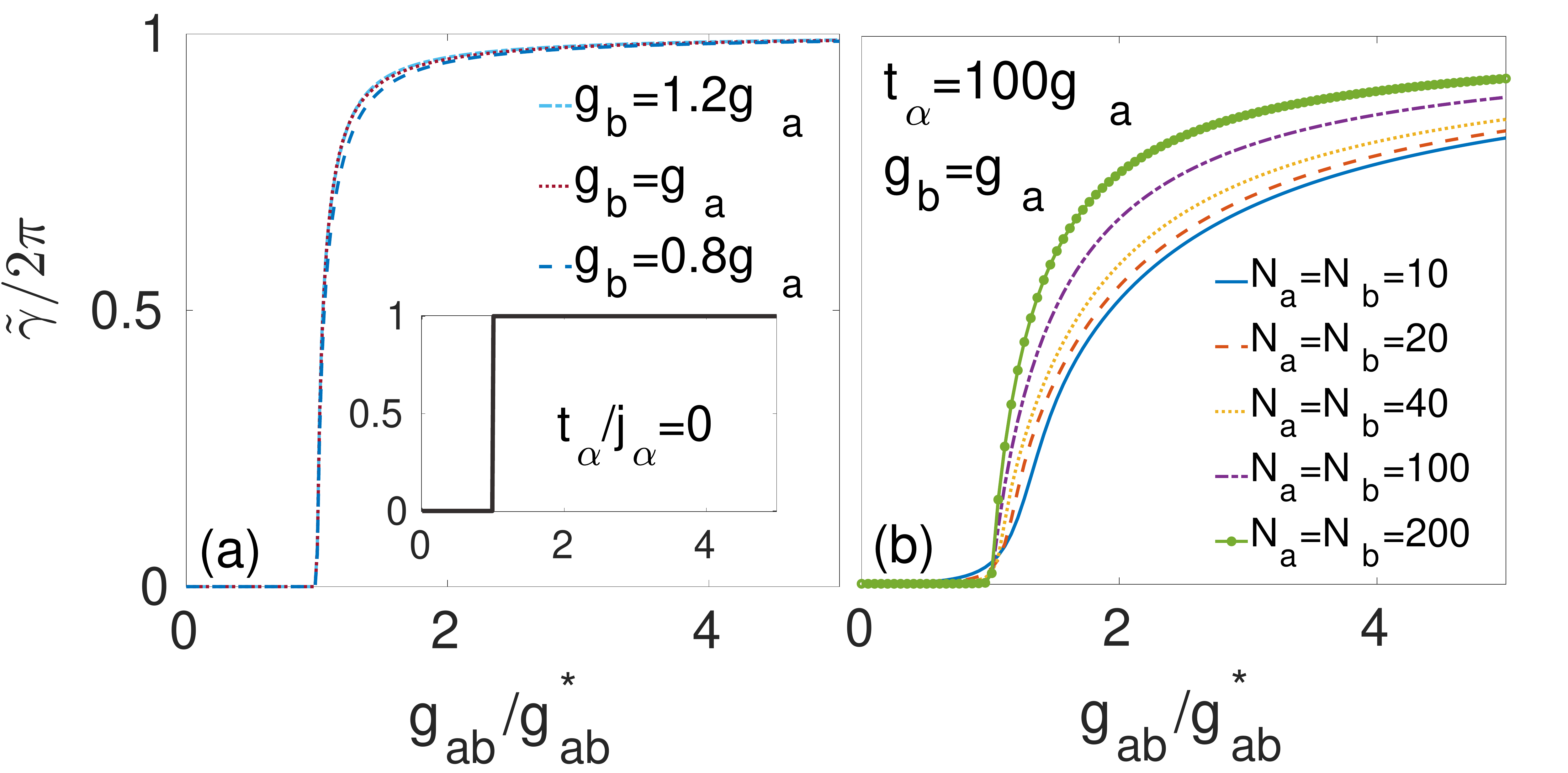}
\caption{The scaled Berry phase of the system for (a) large and (b) small number of particles and as a function of coupling strength $g_{ab}$. Here we use equal tunneling amplitudes for (a)  $ {t}_{\alpha}/j_\alpha=0.1{g}_{a}$ and equal number of particles N$_a $=N$_b=10^4$. In the inset, we take $ {t}_{\alpha}/j_\alpha=0 $~(or equivalently $ j_\alpha=\infty $). For finite number of particles (b), we use $ {t}_{\alpha}=100{g}_{a}$ and obtain the results by solving the eigenstates of the Eq.~(\ref{Eq:Ham_Angular}) numerically.}
\label{fig_berry}
\end{center}
\end{figure}

In this section, we demonstrate that by encircling the critical point in parameter space, where the miscibility-immiscibility transition occurs, a non-trivial Berry phase can be obtained for the system considered in this work. Let us start by introducing the collective angular momentum operators after displacement operation is done. In the limit of large number of particles, it can be found as~[see appendix]
\begin{eqnarray}
\hat{J}_{\alpha x}&\cong & \chi_\alpha  \sqrt{ {\rm N}_\alpha \beta_\alpha (1-\beta_\alpha)}+ \frac{1-2\beta_\alpha}{\sqrt{2(1-\beta_\alpha)}}\hat{X}_\alpha, \label{Disp_Angulara}\\
\hat{J}_{\alpha y}& \cong &-\sqrt{ \frac{1-\beta_\alpha}{2} }\hat{P}_{\alpha}, \label{Disp_Angularb}\\
\hat{J}_{\alpha z}&\cong &{\rm N}_{\alpha} (\beta_\alpha-1/2)+2\chi_{\alpha} \sqrt{{\rm N}_{\alpha} \beta_\alpha} \hat{X}_{\alpha} \nonumber \\ 
&+& \frac{1}{2} (\hat{P}_{\alpha}^2+\hat{X}^2_{\alpha}-1)  \label{Disp_Angularc},
\end{eqnarray}
where $   \chi_{a}= 1$,  $\chi_{b}= -1$, $ \hat{X}_{\alpha} $ and $ \hat{P}_{\alpha} $ are given in Eqs.~(\ref{Eq-posx})-(\ref{Eq-posy}). Here we consider the terms up to 1/N order in the expansion. In the ground state, $ \langle \hat{J}_{\alpha y}\rangle =0$ and  the main contribution to the expectation values of the $  \hat{J}_{\alpha x}$ and $ \hat{J}_{\alpha z} $ comes from the first terms in Eq.(\ref{Disp_Angulara}) and~Eq.(\ref{Disp_Angularc}) respectively. Thus, we can safely neglect the other terms in the thermodynamic limit and obtain 
\begin{equation}
\frac{\langle \hat{J}_{\alpha z}\rangle}{{\rm {N}_\alpha}}=\begin{cases}
-0.5, &  g_{ab}\leq g_{ab}^*\\
(\beta_\alpha-0.5), & g_{ab}> g_{ab}^*
\end{cases}
\label{Eq-Jz}
\end{equation}
and 
\begin{equation}
\frac{\langle \hat{J}_{\alpha x} \rangle }{\sqrt{\rm {N}_\alpha}}=\begin{cases}
0, &  g_{ab}\leq g_{ab}^*\\
\chi_\alpha \sqrt{ \beta_\alpha (1-\beta_\alpha)}, & g_{ab}> g_{ab}^*
\end{cases}
\end{equation}
in which one can clearly observe that above $ g_{ab}^* $ there is a macroscopic excitation for each one. 
If we  introduce a time-dependent unitary transformation $ U(\phi(t))=e^{-i \phi(t) \hat{J}_z} $,  where $ \hat{J}_z=\hat{J}_{az}+\hat{J}_{b z} $ and $ \phi(t) $ is slowly time-varying parameter, which can be defined in an experiment by constructing an adiabatic loop for the interatomic interaction strength between two species via Feshbach resonances~\cite{PhysRevA.82.033609,PhysRevLett.101.040402}. When this phase, $\phi(t)  $, is varied between 0 and 2$ \pi $, a state in phase space will encircle the critical point. Then, Berry phase can be defined in the ground state as~\cite{plastina2006scaling} 
\begin{equation}
\gamma=i \int_0^{2\pi} d\phi \langle \psi| U^{\dagger}(\phi)\frac{d}{d\phi}  U(\phi)| \psi\rangle=2\pi  \langle \hat{J}_z \rangle,
\label{Eq-BP}
\end{equation}
where $ | \psi\rangle $ is the  time-independent  ground-state  wave-function. If we insert Eq.(\ref{Eq-Jz}) into Eq.(\ref{Eq-BP}), the total scaled Berry phase of the system can be defined as  
\begin{equation}
\frac{\tilde{\gamma}}{2\pi }=\begin{cases}
0, &  g_{ab}\leq g_{ab}^*\\
\beta_a+\beta_b, & g_{ab}> g_{ab}^*
\end{cases}
\end{equation}
where we use $ \tilde{\gamma}=1+{\gamma}/{\rm N} $ with N$_a $ = N$_b$  = N. In FIG.~\ref{fig_berry}, we demonstrate the scaled Berry phase of the system as a function of coupling strength, $g_{ab}$, for a finite and infinite number of particles. As it is shown, in the large particle limit, the scaled Berry phase has a zero value for  $g_{ab} \leq g_{ab}^*$ and above $ g_{ab}^* $ increases with increasing the coupling strength and its derivative becomes discontinuous at the critical value, $ g_{ab}^*$. Interestingly, in the thermodynamic limit, there is a step-like transition, which can be seen in the inset of the FIG.~\ref{fig_berry} (a). This is due to solutions of the Eqs.~(\ref{a}) and~(\ref{b}). When $t_\alpha/j_\alpha =0 $, one can see that there is a single solution for each displacement, i.e. $ \beta_\alpha=0.5 $.

It is also possible to obtain non-trivial Berry phase for each species, if we define unitary transformation as $ U(\phi(t))=e^{-i \phi(t) \hat{J}_{\alpha z}} $ and one can obtain $ \gamma_\alpha=2\pi  \langle \hat{J}_{\alpha z} \rangle $. It can be read from Eq.~(\ref{Eq-Jz}) that above the critical coupling strength, $ g_{ab}^*$, there is a finite atomic inversion for each species. This  illustrates the fact that each species has also non-trivial Berry phase.

As it is shown above, increasing the number of particles creates a sharper transition. And/or increasing interspecies coupling ends up with the higher value of the Berry phase. This scenario, however, is not the same for bipartite entanglement. In the following section, we discuss this in more detail.
\section{Bipartite entanglement}\label{sec-Ent}
The detection of the entanglement in an ensemble system by accessing the individual particles is not practical. To obtain a solution in such structures, the global parameters, such as total spin, are used instead~\cite{EsslingerNature2010Dicke,riedel2010atom}. By having such observables in a system, it becomes possible to quantify entanglement~\cite{toth2009spin,sorensen2001Nature,PolzikNature2001ensemble,duan2011many_particle_entanglement,tasgin2017many}. For example, in a recent experiments~\cite{fadel2018spatial,kunkel2018spatially,lange2018entanglement}, bipartite entanglement has been reported for ultracold atomic BECs by measuring collective spins.

 There are several practical criteria~\cite{sorensen2001Nature,HZPRL2006,giovannetti2003characterizing} for the detection of entanglement. These methods are, in general, sufficient, but not necessary. Depending on the structure in a given system, some criteria work better. For instance, in our recent work~\cite{gunay2019entanglement}, we compare different types of the criteria when the system exhibits quantum phase transition, where we observed that criteria, based on bilinear products of spins, can witness the entanglement of two strongly interacting ensembles for small number of particles, in which the criteria of first-order spin fails to detect.
 
In this section, we anticipate the spin form of the criterion derived by  Hillery $ \& $ Zubairy~\cite{HZPRL2006} to investigate bipartite entanglement.
We first introduce the inequality for the detection of entanglement based upon the effective local spin operators~(\ref{Disp_Angulara})-(\ref{Disp_Angularc}), which is given by ~\cite{PhysRevA.86.023626}
\begin{equation}
E_{_{\rm HZ}}= \langle \hat{J}^+_a \hat{J}^-_a \hat{J}^+_b \hat{J}^-_b\rangle-|\langle \hat{J}^+_a\hat{J}^-_b\rangle|^2<0, \label{Eq-HZ}
\end{equation}
where $ E_{_{\rm HZ}} <0$  witnesses the presence of entanglement between the two collective spins. It is important to note that a complete analysis of the derivation of the criterion is beyond the scope of the present paper, and we address the reader to the Refs.~\cite{PhysRevA.86.023626,PhysRevA.80.052335,PhysRevLett.97.170405} for more details. 

\begin{figure}
\begin{center}
\includegraphics[width=86mm,height=45mm]{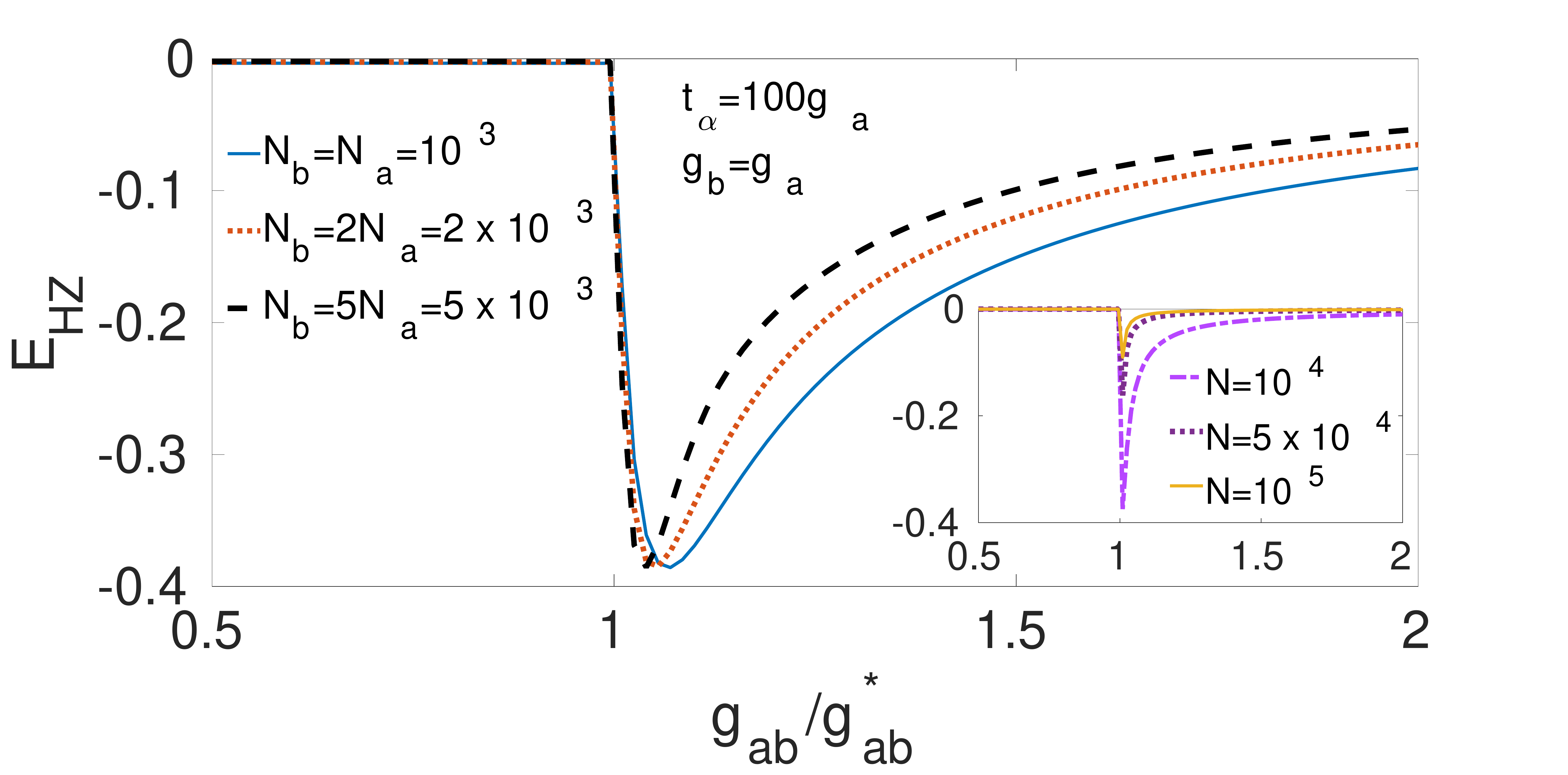}
\caption{Hillery-Zubairy entanglement criterion [Eq.~(\ref{Eq-HZ})] as a function of coupling strength $g_{ab}$. $ E_{_{\rm HZ}} <0$ witnesses the presence of entanglement between the two collective spins. Here we use $ {t}_{\alpha}=100{g}_{a}$ and ${g}_{b}$=${g}_{a}$. In the inset, we take N$_a=$N$_b=$N, and observe that entanglement decays faster as the N increases.}
\label{fig_HZ}
\end{center}
\end{figure}

In FIG.~\ref{fig_HZ}, we demonstrate the results of Eq.~(\ref{Eq-HZ}) as a function of interspecies coupling strength $ g_{ab} $, and for various number of particles.  When the interspecies coupling strength exceeds the critical value there appears a transition also in the entanglement. Unlike the Berry phase, entanglement decays at larger values of the $ g_{ab} $ and/or N. This is due to effective coupling strength, $ \lambda $, in Eq.~(\ref{Bosonic_Ham_BEC_2}), which can be considered as the source of bipartite entanglement. The explanation of this behaviour can be done as follows. When the interatomic coupling strength increases, $ g_{ab} \gg g_{ab}^* $, the value of the displacement parameters approaches to the single value, i.e. $ \beta_\alpha \to 0.5 $, where effective coupling vanishes~[see Eq.~(\ref{Eq:eff_coup})]. The similar story is valid for the increasing number of particles. As, N$ \to \infty $, $ \beta_\alpha \to 0.5 $ [see FIG.~\ref{fig1}]. This can be observed in the inset of the FIG.~\ref{fig_HZ}.
  
In Ref.~\cite{PhysRevA.71.013601}, by using the criterion based on quadrature squeezing~\cite{DGCZ_PRL2000}, it was shown that two-mode entanglement can be present in the stable region~($g_{ab}<g_{ab}^*$). They observed that higher  entanglement can  be  obtained  as  the  system becomes  closer to the critical point. Here, however, $ E_{_{\rm HZ}} $ starts to capture entanglement around the critical value, ~$g_{ab}\gtrsim g_{ab}^*$~[see FIG.~\ref{fig_HZ}]. This shows that the criteria obtained from the squeezing of the spin noise and the ones via squeezing of the bilinear products of the spin are successful in different inseparability regimes~\cite{tasgin2017many}. The experimental testing of these criteria, however, is possible in a system of the two-component BEC by adiabatically changing the interatomic interaction strength around the critical value.

 \section{Summary}\label{sec-sum}
In summary, we have investigated theoretically the ground state properties of the two-component Bose-Einstein-condensate trapped in a double-well potential. We observe that the system can undergo a QPT at a critical coupling strength.  We obtain the effective Hamiltonians and associated ground state wavefunctions to describe the system in each of its mixed and separated phases. The non-trivial geometric phase is found near the critical coupling in the limit of small and large numbers of particles, where we observe a step-like transition in the thermodynamic limit. The accuracy of the model is confirmed by comparison with the exact numerical solution in the small number of particles limit~[see Figure~\ref{fig_berry}]. We also anticipate the spin form of Hillery $\&$ Zubairy criterion to quantify entanglement across a QPT. It is observed that entanglement witness, $ E_{_{\rm HZ}}  $, decays with increasing interspecies interaction strength and/or number of particles. 
The tunable interactions between the two species via Feshbach resonances, making the model a promising
simulator for this kind of structures, and can find potential in the area of quantum communication~\cite{BraunsteinNaturePhot2015} and quantum sensing~\cite{degen2017quantum}.

\section*{Acknowledgments}
 This research was supported by  The Scientific and Technological Research Council of Turkey (TUBITAK) Grant No. 117F118. 
\appendix
\renewcommand{\thesubsection}{\Alph{subsection}}
\section*{Appendix}
Here we show the derivations of Eqs.~(\ref{Disp_Angulara}) and (\ref{Disp_Angularb}). By inserting displaced operators defined in Eq.~(\ref{Displacement}) into the Eq.~(\ref{Eq:Hols_Prim}), it becomes: 
\begin{eqnarray}
\hat{J}_{\alpha}^{+}&=&(\hat{d}_{\alpha}^{\dagger} +\chi_\alpha \sqrt{{\rm N}_\alpha \beta_\alpha}) \sqrt{1-\beta_\alpha}\sqrt{1-\xi_\alpha}, \label{Eq:xi}\\
\xi_\alpha &\equiv &\bigg(1-\frac{\hat{d}_{\alpha}^{\dagger}\hat{d}_{\alpha}+\chi_\alpha \sqrt{{\rm N}_\alpha \beta_\alpha}}{{\rm N}_\alpha (1-\beta_\alpha)}\bigg).
\end{eqnarray}
After expanding the last term, $ \sqrt{1-\xi_\alpha} $, in Eq.~(\ref{Eq:xi}) up to 1/N order, one can arrive:
\begin{eqnarray}
\hat{J}_{\alpha}^{+}&\approx & \sqrt{1-\beta_\alpha} \bigg[ (\hat{d}_{\alpha}^{\dagger} +\sqrt{{\rm N}_1 \beta_\alpha})-\frac{{\beta_\alpha} }{{1-\beta_\alpha}} \frac{\hat{X}_{\alpha}}{\sqrt{2}}\bigg].
\end{eqnarray}
Similarly, one can derive lowering component, $\hat{J}_{\alpha}^{-} =( \hat{J}_{\alpha}^{+})^\dagger$. By using the definitions: $ \hat{J}_{\alpha x}=({\hat{J}_{\alpha}^{+}+\hat{J}_{\alpha}^{-}})/{2} $ and $ \hat{J}_{\alpha y}=({\hat{J}_{\alpha}^{+}-\hat{J}_{\alpha}^{-}})/{2i} $, we derive Eqs.~(\ref{Disp_Angulara}) and (\ref{Disp_Angularb}) as
\begin{eqnarray}
\hat{J}_{\alpha x}&\cong & \chi_\alpha  \sqrt{ {\rm N}_\alpha \beta_\alpha (1-\beta_\alpha)}+ \frac{1-2\beta_\alpha}{\sqrt{2(1-\beta_\alpha)}}\hat{X}_\alpha, \label{app:Disp_Angulara}\\
\hat{J}_{\alpha y}& \cong &-\sqrt{ \frac{1-\beta_\alpha}{2} }\hat{P}_{\alpha},
\end{eqnarray}
It is straightforward to obtain $ z $-component of the angular momentum operator, by inserting Eqs.~(\ref{Displacement}), (\ref{Eq-posx}) and (\ref{Eq-posy}) into the definition of $ \hat{J}_{\alpha z}  $ given in Eq.~(\ref{Eq:Hols_Prim}).
\bibliography{bibliography}

\begin{thebibliography}{64}%
\makeatletter
\providecommand \@ifxundefined [1]{%
 \@ifx{#1\undefined}
}%
\providecommand \@ifnum [1]{%
 \ifnum #1\expandafter \@firstoftwo
 \else \expandafter \@secondoftwo
 \fi
}%
\providecommand \@ifx [1]{%
 \ifx #1\expandafter \@firstoftwo
 \else \expandafter \@secondoftwo
 \fi
}%
\providecommand \natexlab [1]{#1}%
\providecommand \enquote  [1]{``#1''}%
\providecommand \bibnamefont  [1]{#1}%
\providecommand \bibfnamefont [1]{#1}%
\providecommand \citenamefont [1]{#1}%
\providecommand \href@noop [0]{\@secondoftwo}%
\providecommand \href [0]{\begingroup \@sanitize@url \@href}%
\providecommand \@href[1]{\@@startlink{#1}\@@href}%
\providecommand \@@href[1]{\endgroup#1\@@endlink}%
\providecommand \@sanitize@url [0]{\catcode `\\12\catcode `\$12\catcode
  `\&12\catcode `\#12\catcode `\^12\catcode `\_12\catcode `\%12\relax}%
\providecommand \@@startlink[1]{}%
\providecommand \@@endlink[0]{}%
\providecommand \url  [0]{\begingroup\@sanitize@url \@url }%
\providecommand \@url [1]{\endgroup\@href {#1}{\urlprefix }}%
\providecommand \urlprefix  [0]{URL }%
\providecommand \Eprint [0]{\href }%
\providecommand \doibase [0]{http://dx.doi.org/}%
\providecommand \selectlanguage [0]{\@gobble}%
\providecommand \bibinfo  [0]{\@secondoftwo}%
\providecommand \bibfield  [0]{\@secondoftwo}%
\providecommand \translation [1]{[#1]}%
\providecommand \BibitemOpen [0]{}%
\providecommand \bibitemStop [0]{}%
\providecommand \bibitemNoStop [0]{.\EOS\space}%
\providecommand \EOS [0]{\spacefactor3000\relax}%
\providecommand \BibitemShut  [1]{\csname bibitem#1\endcsname}%
\let\auto@bib@innerbib\@empty
\bibitem [{\citenamefont {Qiu}\ \emph {et~al.}(2010)\citenamefont {Qiu},
  \citenamefont {Tian},\ and\ \citenamefont {Fu}}]{PhysRevA.81.043613}%
  \BibitemOpen
  \bibfield  {author} {\bibinfo {author} {\bibfnamefont {Haibo}\ \bibnamefont
  {Qiu}}, \bibinfo {author} {\bibfnamefont {Jing}\ \bibnamefont {Tian}}, \ and\
  \bibinfo {author} {\bibfnamefont {Li-Bin}\ \bibnamefont {Fu}},\ }\bibfield
  {title} {\enquote {\bibinfo {title} {{Collective dynamics of two-species
  Bose-Einstein-condensate mixtures in a double-well potential}},}\ }\href
  {\doibase 10.1103/PhysRevA.81.043613} {\bibfield  {journal} {\bibinfo
  {journal} {Phys. Rev. A}\ }\textbf {\bibinfo {volume} {81}},\ \bibinfo
  {pages} {043613} (\bibinfo {year} {2010})}\BibitemShut {NoStop}%
\bibitem [{\citenamefont {Lee}\ \emph {et~al.}(2016)\citenamefont {Lee},
  \citenamefont {J\o{}rgensen}, \citenamefont {Liu}, \citenamefont {Wacker},
  \citenamefont {Arlt},\ and\ \citenamefont {Proukakis}}]{PhysRevA.94.013602}%
  \BibitemOpen
  \bibfield  {author} {\bibinfo {author} {\bibfnamefont {Kean~Loon}\
  \bibnamefont {Lee}}, \bibinfo {author} {\bibfnamefont {Nils~B.}\ \bibnamefont
  {J\o{}rgensen}}, \bibinfo {author} {\bibfnamefont {I-Kang}\ \bibnamefont
  {Liu}}, \bibinfo {author} {\bibfnamefont {Lars}\ \bibnamefont {Wacker}},
  \bibinfo {author} {\bibfnamefont {Jan~J.}\ \bibnamefont {Arlt}}, \ and\
  \bibinfo {author} {\bibfnamefont {Nick~P.}\ \bibnamefont {Proukakis}},\
  }\bibfield  {title} {\enquote {\bibinfo {title} {{Phase separation and
  dynamics of two-component Bose-Einstein condensates}},}\ }\href {\doibase
  10.1103/PhysRevA.94.013602} {\bibfield  {journal} {\bibinfo  {journal} {Phys.
  Rev. A}\ }\textbf {\bibinfo {volume} {94}},\ \bibinfo {pages} {013602}
  (\bibinfo {year} {2016})}\BibitemShut {NoStop}%
\bibitem [{\citenamefont {Mertes}\ \emph {et~al.}(2007)\citenamefont {Mertes},
  \citenamefont {Merrill}, \citenamefont {Carretero-Gonz{\'a}lez},
  \citenamefont {Frantzeskakis}, \citenamefont {Kevrekidis},\ and\
  \citenamefont {Hall}}]{mertes2007nonequilibrium}%
  \BibitemOpen
  \bibfield  {author} {\bibinfo {author} {\bibfnamefont {KM}~\bibnamefont
  {Mertes}}, \bibinfo {author} {\bibfnamefont {JW}~\bibnamefont {Merrill}},
  \bibinfo {author} {\bibfnamefont {R}~\bibnamefont {Carretero-Gonz{\'a}lez}},
  \bibinfo {author} {\bibfnamefont {DJ}~\bibnamefont {Frantzeskakis}}, \bibinfo
  {author} {\bibfnamefont {PG}~\bibnamefont {Kevrekidis}}, \ and\ \bibinfo
  {author} {\bibfnamefont {DS}~\bibnamefont {Hall}},\ }\bibfield  {title}
  {\enquote {\bibinfo {title} {{Nonequilibrium dynamics and superfluid ring
  excitations in binary Bose-Einstein condensates}},}\ }\href@noop {}
  {\bibfield  {journal} {\bibinfo  {journal} {Phys. Rev.Lett.}\ }\textbf
  {\bibinfo {volume} {99}},\ \bibinfo {pages} {190402} (\bibinfo {year}
  {2007})}\BibitemShut {NoStop}%
\bibitem [{\citenamefont {Molony}\ \emph {et~al.}(2014)\citenamefont {Molony},
  \citenamefont {Gregory}, \citenamefont {Ji}, \citenamefont {Lu},
  \citenamefont {K\"oppinger}, \citenamefont {Le~Sueur}, \citenamefont
  {Blackley}, \citenamefont {Hutson},\ and\ \citenamefont
  {Cornish}}]{PhysRevLett.113.255301}%
  \BibitemOpen
  \bibfield  {author} {\bibinfo {author} {\bibfnamefont {Peter~K.}\
  \bibnamefont {Molony}}, \bibinfo {author} {\bibfnamefont {Philip~D.}\
  \bibnamefont {Gregory}}, \bibinfo {author} {\bibfnamefont {Zhonghua}\
  \bibnamefont {Ji}}, \bibinfo {author} {\bibfnamefont {Bo}~\bibnamefont {Lu}},
  \bibinfo {author} {\bibfnamefont {Michael~P.}\ \bibnamefont {K\"oppinger}},
  \bibinfo {author} {\bibfnamefont {C.~Ruth}\ \bibnamefont {Le~Sueur}},
  \bibinfo {author} {\bibfnamefont {Caroline~L.}\ \bibnamefont {Blackley}},
  \bibinfo {author} {\bibfnamefont {Jeremy~M.}\ \bibnamefont {Hutson}}, \ and\
  \bibinfo {author} {\bibfnamefont {Simon~L.}\ \bibnamefont {Cornish}},\
  }\bibfield  {title} {\enquote {\bibinfo {title} {Creation of ultracold
  $^{87}\mathrm{Rb}^{133}$ - $\mathrm{Cs}$ molecules in the rovibrational
  ground state},}\ }\href {\doibase 10.1103/PhysRevLett.113.255301} {\bibfield
  {journal} {\bibinfo  {journal} {Phys. Rev. Lett.}\ }\textbf {\bibinfo
  {volume} {113}},\ \bibinfo {pages} {255301} (\bibinfo {year}
  {2014})}\BibitemShut {NoStop}%
\bibitem [{\citenamefont {Ng}\ and\ \citenamefont
  {Leung}(2005)}]{PhysRevA.71.013601}%
  \BibitemOpen
  \bibfield  {author} {\bibinfo {author} {\bibfnamefont {H.~T.}\ \bibnamefont
  {Ng}}\ and\ \bibinfo {author} {\bibfnamefont {P.~T.}\ \bibnamefont {Leung}},\
  }\bibfield  {title} {\enquote {\bibinfo {title} {{Two-mode entanglement in
  two-component Bose-Einstein condensates}},}\ }\href {\doibase
  10.1103/PhysRevA.71.013601} {\bibfield  {journal} {\bibinfo  {journal} {Phys.
  Rev. A}\ }\textbf {\bibinfo {volume} {71}},\ \bibinfo {pages} {013601}
  (\bibinfo {year} {2005})}\BibitemShut {NoStop}%
\bibitem [{\citenamefont {He}\ \emph {et~al.}(2011)\citenamefont {He},
  \citenamefont {Reid}, \citenamefont {Vaughan}, \citenamefont {Gross},
  \citenamefont {Oberthaler},\ and\ \citenamefont
  {Drummond}}]{PhysRevLett.106.120405}%
  \BibitemOpen
  \bibfield  {author} {\bibinfo {author} {\bibfnamefont {Q.~Y.}\ \bibnamefont
  {He}}, \bibinfo {author} {\bibfnamefont {M.~D.}\ \bibnamefont {Reid}},
  \bibinfo {author} {\bibfnamefont {T.~G.}\ \bibnamefont {Vaughan}}, \bibinfo
  {author} {\bibfnamefont {C.}~\bibnamefont {Gross}}, \bibinfo {author}
  {\bibfnamefont {M.}~\bibnamefont {Oberthaler}}, \ and\ \bibinfo {author}
  {\bibfnamefont {P.~D.}\ \bibnamefont {Drummond}},\ }\bibfield  {title}
  {\enquote {\bibinfo {title} {{Einstein-Podolsky-Rosen entanglement strategies
  in two-well Bose-Einstein condensates}},}\ }\href {\doibase
  10.1103/PhysRevLett.106.120405} {\bibfield  {journal} {\bibinfo  {journal}
  {Phys. Rev. Lett.}\ }\textbf {\bibinfo {volume} {106}},\ \bibinfo {pages}
  {120405} (\bibinfo {year} {2011})}\BibitemShut {NoStop}%
\bibitem [{\citenamefont {Levy}\ \emph {et~al.}(2007)\citenamefont {Levy},
  \citenamefont {Lahoud}, \citenamefont {Shomroni},\ and\ \citenamefont
  {Steinhauer}}]{levy2007ac}%
  \BibitemOpen
  \bibfield  {author} {\bibinfo {author} {\bibfnamefont {S}~\bibnamefont
  {Levy}}, \bibinfo {author} {\bibfnamefont {E}~\bibnamefont {Lahoud}},
  \bibinfo {author} {\bibfnamefont {I}~\bibnamefont {Shomroni}}, \ and\
  \bibinfo {author} {\bibfnamefont {J}~\bibnamefont {Steinhauer}},\ }\bibfield
  {title} {\enquote {\bibinfo {title} {{The ac and dc Josephson effects in a
  Bose-Einstein condensate}},}\ }\href@noop {} {\bibfield  {journal} {\bibinfo
  {journal} {Nature}\ }\textbf {\bibinfo {volume} {449}},\ \bibinfo {pages}
  {579} (\bibinfo {year} {2007})}\BibitemShut {NoStop}%
\bibitem [{\citenamefont {Zibold}\ \emph {et~al.}(2010)\citenamefont {Zibold},
  \citenamefont {Nicklas}, \citenamefont {Gross},\ and\ \citenamefont
  {Oberthaler}}]{zibold2010classical}%
  \BibitemOpen
  \bibfield  {author} {\bibinfo {author} {\bibfnamefont {Tilman}\ \bibnamefont
  {Zibold}}, \bibinfo {author} {\bibfnamefont {Eike}\ \bibnamefont {Nicklas}},
  \bibinfo {author} {\bibfnamefont {Christian}\ \bibnamefont {Gross}}, \ and\
  \bibinfo {author} {\bibfnamefont {Markus~K}\ \bibnamefont {Oberthaler}},\
  }\bibfield  {title} {\enquote {\bibinfo {title} {{Classical bifurcation at
  the transition from Rabi to Josephson dynamics}},}\ }\href@noop {} {\bibfield
   {journal} {\bibinfo  {journal} {Phys. Rev. Lett.}\ }\textbf {\bibinfo
  {volume} {105}},\ \bibinfo {pages} {204101} (\bibinfo {year}
  {2010})}\BibitemShut {NoStop}%
\bibitem [{\citenamefont {Walters}\ and\ \citenamefont
  {Fairbank}(1956)}]{PhysRev.103.262.2}%
  \BibitemOpen
  \bibfield  {author} {\bibinfo {author} {\bibfnamefont {G.~K.}\ \bibnamefont
  {Walters}}\ and\ \bibinfo {author} {\bibfnamefont {W.~M.}\ \bibnamefont
  {Fairbank}},\ }\bibfield  {title} {\enquote {\bibinfo {title} {Phase
  separation in ${^{3}{He}}$ - ${^{4}{He}}$ solutions},}\ }\href {\doibase
  10.1103/PhysRev.103.262.2} {\bibfield  {journal} {\bibinfo  {journal} {Phys.
  Rev.}\ }\textbf {\bibinfo {volume} {103}},\ \bibinfo {pages} {262--263}
  (\bibinfo {year} {1956})}\BibitemShut {NoStop}%
\bibitem [{\citenamefont {Myatt}\ \emph {et~al.}(1997)\citenamefont {Myatt},
  \citenamefont {Burt}, \citenamefont {Ghrist}, \citenamefont {Cornell},\ and\
  \citenamefont {Wieman}}]{myatt1997production}%
  \BibitemOpen
  \bibfield  {author} {\bibinfo {author} {\bibfnamefont {CJ}~\bibnamefont
  {Myatt}}, \bibinfo {author} {\bibfnamefont {EA}~\bibnamefont {Burt}},
  \bibinfo {author} {\bibfnamefont {RW}~\bibnamefont {Ghrist}}, \bibinfo
  {author} {\bibfnamefont {Eric~A}\ \bibnamefont {Cornell}}, \ and\ \bibinfo
  {author} {\bibfnamefont {CE}~\bibnamefont {Wieman}},\ }\bibfield  {title}
  {\enquote {\bibinfo {title} {{Production of two overlapping Bose-Einstein
  condensates by sympathetic cooling}},}\ }\href@noop {} {\bibfield  {journal}
  {\bibinfo  {journal} {Phys. Rev. Lett.}\ }\textbf {\bibinfo {volume} {78}},\
  \bibinfo {pages} {586} (\bibinfo {year} {1997})}\BibitemShut {NoStop}%
\bibitem [{\citenamefont {Hall}\ \emph {et~al.}(1998)\citenamefont {Hall},
  \citenamefont {Matthews}, \citenamefont {Ensher}, \citenamefont {Wieman},\
  and\ \citenamefont {Cornell}}]{PhysRevLett.81.1539}%
  \BibitemOpen
  \bibfield  {author} {\bibinfo {author} {\bibfnamefont {D.~S.}\ \bibnamefont
  {Hall}}, \bibinfo {author} {\bibfnamefont {M.~R.}\ \bibnamefont {Matthews}},
  \bibinfo {author} {\bibfnamefont {J.~R.}\ \bibnamefont {Ensher}}, \bibinfo
  {author} {\bibfnamefont {C.~E.}\ \bibnamefont {Wieman}}, \ and\ \bibinfo
  {author} {\bibfnamefont {E.~A.}\ \bibnamefont {Cornell}},\ }\bibfield
  {title} {\enquote {\bibinfo {title} {{Dynamics of Component Separation in a
  Binary Mixture of Bose-Einstein Condensates}},}\ }\href {\doibase
  10.1103/PhysRevLett.81.1539} {\bibfield  {journal} {\bibinfo  {journal}
  {Phys. Rev. Lett.}\ }\textbf {\bibinfo {volume} {81}},\ \bibinfo {pages}
  {1539--1542} (\bibinfo {year} {1998})}\BibitemShut {NoStop}%
\bibitem [{\citenamefont {Tojo}\ \emph {et~al.}(2010)\citenamefont {Tojo},
  \citenamefont {Taguchi}, \citenamefont {Masuyama}, \citenamefont {Hayashi},
  \citenamefont {Saito},\ and\ \citenamefont {Hirano}}]{PhysRevA.82.033609}%
  \BibitemOpen
  \bibfield  {author} {\bibinfo {author} {\bibfnamefont {Satoshi}\ \bibnamefont
  {Tojo}}, \bibinfo {author} {\bibfnamefont {Yoshihisa}\ \bibnamefont
  {Taguchi}}, \bibinfo {author} {\bibfnamefont {Yuta}\ \bibnamefont
  {Masuyama}}, \bibinfo {author} {\bibfnamefont {Taro}\ \bibnamefont
  {Hayashi}}, \bibinfo {author} {\bibfnamefont {Hiroki}\ \bibnamefont {Saito}},
  \ and\ \bibinfo {author} {\bibfnamefont {Takuya}\ \bibnamefont {Hirano}},\
  }\bibfield  {title} {\enquote {\bibinfo {title} {{Controlling phase
  separation of binary Bose-Einstein condensates via mixed-spin-channel
  Feshbach resonance}},}\ }\href {\doibase 10.1103/PhysRevA.82.033609}
  {\bibfield  {journal} {\bibinfo  {journal} {Phys. Rev. A}\ }\textbf {\bibinfo
  {volume} {82}},\ \bibinfo {pages} {033609} (\bibinfo {year}
  {2010})}\BibitemShut {NoStop}%
\bibitem [{\citenamefont {Papp}\ \emph {et~al.}(2008)\citenamefont {Papp},
  \citenamefont {Pino},\ and\ \citenamefont {Wieman}}]{PhysRevLett.101.040402}%
  \BibitemOpen
  \bibfield  {author} {\bibinfo {author} {\bibfnamefont {S.~B.}\ \bibnamefont
  {Papp}}, \bibinfo {author} {\bibfnamefont {J.~M.}\ \bibnamefont {Pino}}, \
  and\ \bibinfo {author} {\bibfnamefont {C.~E.}\ \bibnamefont {Wieman}},\
  }\bibfield  {title} {\enquote {\bibinfo {title} {{Tunable Miscibility in a
  Dual-Species Bose-Einstein Condensate}},}\ }\href {\doibase
  10.1103/PhysRevLett.101.040402} {\bibfield  {journal} {\bibinfo  {journal}
  {Phys. Rev. Lett.}\ }\textbf {\bibinfo {volume} {101}},\ \bibinfo {pages}
  {040402} (\bibinfo {year} {2008})}\BibitemShut {NoStop}%
\bibitem [{\citenamefont {Simoni}\ \emph {et~al.}(2003)\citenamefont {Simoni},
  \citenamefont {Ferlaino}, \citenamefont {Roati}, \citenamefont {Modugno},\
  and\ \citenamefont {Inguscio}}]{simoni2003magnetic}%
  \BibitemOpen
  \bibfield  {author} {\bibinfo {author} {\bibfnamefont {Andrea}\ \bibnamefont
  {Simoni}}, \bibinfo {author} {\bibfnamefont {Francesca}\ \bibnamefont
  {Ferlaino}}, \bibinfo {author} {\bibfnamefont {Giacomo}\ \bibnamefont
  {Roati}}, \bibinfo {author} {\bibfnamefont {Giovanni}\ \bibnamefont
  {Modugno}}, \ and\ \bibinfo {author} {\bibfnamefont {Massimo}\ \bibnamefont
  {Inguscio}},\ }\bibfield  {title} {\enquote {\bibinfo {title} {{Magnetic
  control of the interaction in ultracold K-Rb mixtures}},}\ }\href@noop {}
  {\bibfield  {journal} {\bibinfo  {journal} {Phys. Rev. Lett.}\ }\textbf
  {\bibinfo {volume} {90}},\ \bibinfo {pages} {163202} (\bibinfo {year}
  {2003})}\BibitemShut {NoStop}%
\bibitem [{\citenamefont {Roy}\ \emph {et~al.}(2017)\citenamefont {Roy},
  \citenamefont {Green}, \citenamefont {Bowler},\ and\ \citenamefont
  {Gupta}}]{roy2017two}%
  \BibitemOpen
  \bibfield  {author} {\bibinfo {author} {\bibfnamefont {Richard}\ \bibnamefont
  {Roy}}, \bibinfo {author} {\bibfnamefont {Alaina}\ \bibnamefont {Green}},
  \bibinfo {author} {\bibfnamefont {Ryan}\ \bibnamefont {Bowler}}, \ and\
  \bibinfo {author} {\bibfnamefont {Subhadeep}\ \bibnamefont {Gupta}},\
  }\bibfield  {title} {\enquote {\bibinfo {title} {{Two-element mixture of Bose
  and Fermi superfluids}},}\ }\href@noop {} {\bibfield  {journal} {\bibinfo
  {journal} {Phys. Rev. Lett.}\ }\textbf {\bibinfo {volume} {118}},\ \bibinfo
  {pages} {055301} (\bibinfo {year} {2017})}\BibitemShut {NoStop}%
\bibitem [{\citenamefont {McCarron}\ \emph {et~al.}(2011)\citenamefont
  {McCarron}, \citenamefont {Cho}, \citenamefont {Jenkin}, \citenamefont
  {K\"oppinger},\ and\ \citenamefont {Cornish}}]{PhysRevA.84.011603}%
  \BibitemOpen
  \bibfield  {author} {\bibinfo {author} {\bibfnamefont {D.~J.}\ \bibnamefont
  {McCarron}}, \bibinfo {author} {\bibfnamefont {H.~W.}\ \bibnamefont {Cho}},
  \bibinfo {author} {\bibfnamefont {D.~L.}\ \bibnamefont {Jenkin}}, \bibinfo
  {author} {\bibfnamefont {M.~P.}\ \bibnamefont {K\"oppinger}}, \ and\ \bibinfo
  {author} {\bibfnamefont {S.~L.}\ \bibnamefont {Cornish}},\ }\bibfield
  {title} {\enquote {\bibinfo {title} {{Dual-species Bose-Einstein condensate
  of $^{87}\mathrm{Rb}$ and $^{133}\mathrm{Cs}$}},}\ }\href {\doibase
  10.1103/PhysRevA.84.011603} {\bibfield  {journal} {\bibinfo  {journal} {Phys.
  Rev. A}\ }\textbf {\bibinfo {volume} {84}},\ \bibinfo {pages} {011603}
  (\bibinfo {year} {2011})}\BibitemShut {NoStop}%
\bibitem [{\citenamefont {Lercher}\ \emph {et~al.}(2011)\citenamefont
  {Lercher}, \citenamefont {Takekoshi}, \citenamefont {Debatin}, \citenamefont
  {Schuster}, \citenamefont {Rameshan}, \citenamefont {Ferlaino}, \citenamefont
  {Grimm},\ and\ \citenamefont {N{\"a}gerl}}]{lercher2011production}%
  \BibitemOpen
  \bibfield  {author} {\bibinfo {author} {\bibfnamefont {AD}~\bibnamefont
  {Lercher}}, \bibinfo {author} {\bibfnamefont {T}~\bibnamefont {Takekoshi}},
  \bibinfo {author} {\bibfnamefont {M}~\bibnamefont {Debatin}}, \bibinfo
  {author} {\bibfnamefont {B}~\bibnamefont {Schuster}}, \bibinfo {author}
  {\bibfnamefont {R}~\bibnamefont {Rameshan}}, \bibinfo {author} {\bibfnamefont
  {F}~\bibnamefont {Ferlaino}}, \bibinfo {author} {\bibfnamefont
  {R}~\bibnamefont {Grimm}}, \ and\ \bibinfo {author} {\bibfnamefont {H-C}\
  \bibnamefont {N{\"a}gerl}},\ }\bibfield  {title} {\enquote {\bibinfo {title}
  {{Production of a dual-species Bose-Einstein condensate of Rb and Cs
  atoms}},}\ }\href@noop {} {\bibfield  {journal} {\bibinfo  {journal} {The
  European Physical Journal D}\ }\textbf {\bibinfo {volume} {65}},\ \bibinfo
  {pages} {3--9} (\bibinfo {year} {2011})}\BibitemShut {NoStop}%
\bibitem [{\citenamefont {Lingua}\ and\ \citenamefont
  {Penna}(2017)}]{PhysRevE.95.062142}%
  \BibitemOpen
  \bibfield  {author} {\bibinfo {author} {\bibfnamefont {F.}~\bibnamefont
  {Lingua}}\ and\ \bibinfo {author} {\bibfnamefont {V.}~\bibnamefont {Penna}},\
  }\bibfield  {title} {\enquote {\bibinfo {title} {{Continuous-variable
  approach to the spectral properties and quantum states of the two-component
  Bose-Hubbard dimer}},}\ }\href {\doibase 10.1103/PhysRevE.95.062142}
  {\bibfield  {journal} {\bibinfo  {journal} {Phys. Rev. E}\ }\textbf {\bibinfo
  {volume} {95}},\ \bibinfo {pages} {062142} (\bibinfo {year}
  {2017})}\BibitemShut {NoStop}%
\bibitem [{\citenamefont {Smerzi}\ \emph {et~al.}(1997)\citenamefont {Smerzi},
  \citenamefont {Fantoni}, \citenamefont {Giovanazzi},\ and\ \citenamefont
  {Shenoy}}]{PhysRevLett.79.4950}%
  \BibitemOpen
  \bibfield  {author} {\bibinfo {author} {\bibfnamefont {A.}~\bibnamefont
  {Smerzi}}, \bibinfo {author} {\bibfnamefont {S.}~\bibnamefont {Fantoni}},
  \bibinfo {author} {\bibfnamefont {S.}~\bibnamefont {Giovanazzi}}, \ and\
  \bibinfo {author} {\bibfnamefont {S.~R.}\ \bibnamefont {Shenoy}},\ }\bibfield
   {title} {\enquote {\bibinfo {title} {{Quantum coherent atomic tunneling
  between two trapped Bose-Einstein condensates}},}\ }\href {\doibase
  10.1103/PhysRevLett.79.4950} {\bibfield  {journal} {\bibinfo  {journal}
  {Phys. Rev. Lett.}\ }\textbf {\bibinfo {volume} {79}},\ \bibinfo {pages}
  {4950--4953} (\bibinfo {year} {1997})}\BibitemShut {NoStop}%
\bibitem [{\citenamefont {Lingua}\ \emph {et~al.}(2016)\citenamefont {Lingua},
  \citenamefont {Mazzarella},\ and\ \citenamefont
  {Penna}}]{lingua2016delocalization}%
  \BibitemOpen
  \bibfield  {author} {\bibinfo {author} {\bibfnamefont {F}~\bibnamefont
  {Lingua}}, \bibinfo {author} {\bibfnamefont {G}~\bibnamefont {Mazzarella}}, \
  and\ \bibinfo {author} {\bibfnamefont {V}~\bibnamefont {Penna}},\ }\bibfield
  {title} {\enquote {\bibinfo {title} {Delocalization effects, entanglement
  entropy and spectral collapse of boson mixtures in a double well},}\
  }\href@noop {} {\bibfield  {journal} {\bibinfo  {journal} {Journal of Physics
  B: Atomic, Molecular and Optical Physics}\ }\textbf {\bibinfo {volume}
  {49}},\ \bibinfo {pages} {205005} (\bibinfo {year} {2016})}\BibitemShut
  {NoStop}%
\bibitem [{\citenamefont {Lingua}\ \emph {et~al.}(2018)\citenamefont {Lingua},
  \citenamefont {Richaud},\ and\ \citenamefont {Penna}}]{lingua2018residual}%
  \BibitemOpen
  \bibfield  {author} {\bibinfo {author} {\bibfnamefont {Fabio}\ \bibnamefont
  {Lingua}}, \bibinfo {author} {\bibfnamefont {Andrea}\ \bibnamefont
  {Richaud}}, \ and\ \bibinfo {author} {\bibfnamefont {Vittorio}\ \bibnamefont
  {Penna}},\ }\bibfield  {title} {\enquote {\bibinfo {title} {Residual entropy
  and critical behavior of two interacting boson species in a double well},}\
  }\href@noop {} {\bibfield  {journal} {\bibinfo  {journal} {Entropy}\ }\textbf
  {\bibinfo {volume} {20}},\ \bibinfo {pages} {84} (\bibinfo {year}
  {2018})}\BibitemShut {NoStop}%
\bibitem [{\citenamefont {Fu}\ and\ \citenamefont
  {Liu}(2006)}]{PhysRevA.74.063614}%
  \BibitemOpen
  \bibfield  {author} {\bibinfo {author} {\bibfnamefont {L.}~\bibnamefont
  {Fu}}\ and\ \bibinfo {author} {\bibfnamefont {J.}~\bibnamefont {Liu}},\
  }\bibfield  {title} {\enquote {\bibinfo {title} {Quantum entanglement
  manifestation of transition to nonlinear self-trapping for bose-einstein
  condensates in a symmetric double well},}\ }\href {\doibase
  10.1103/PhysRevA.74.063614} {\bibfield  {journal} {\bibinfo  {journal} {Phys.
  Rev. A}\ }\textbf {\bibinfo {volume} {74}},\ \bibinfo {pages} {063614}
  (\bibinfo {year} {2006})}\BibitemShut {NoStop}%
\bibitem [{\citenamefont {Tian}\ and\ \citenamefont
  {Qiu}(2016)}]{tian2016dynamical}%
  \BibitemOpen
  \bibfield  {author} {\bibinfo {author} {\bibfnamefont {Jing}\ \bibnamefont
  {Tian}}\ and\ \bibinfo {author} {\bibfnamefont {Haibo}\ \bibnamefont {Qiu}},\
  }\bibfield  {title} {\enquote {\bibinfo {title} {{A Dynamical Phase
  Transition of Binary Species BECs Mixtures in a Double Well Potential}},}\
  }\href@noop {} {\bibfield  {journal} {\bibinfo  {journal} {International
  Journal of Theoretical Physics}\ }\textbf {\bibinfo {volume} {2}},\ \bibinfo
  {pages} {321--327} (\bibinfo {year} {2016})}\BibitemShut {NoStop}%
\bibitem [{\citenamefont {Ng}\ and\ \citenamefont
  {Chu}(2011)}]{PhysRevA.84.023629}%
  \BibitemOpen
  \bibfield  {author} {\bibinfo {author} {\bibfnamefont {H.~T.}\ \bibnamefont
  {Ng}}\ and\ \bibinfo {author} {\bibfnamefont {Shih-I}\ \bibnamefont {Chu}},\
  }\bibfield  {title} {\enquote {\bibinfo {title} {Steady-state entanglement in
  a double-well bose-einstein condensate through coupling to a superconducting
  resonator},}\ }\href {\doibase 10.1103/PhysRevA.84.023629} {\bibfield
  {journal} {\bibinfo  {journal} {Phys. Rev. A}\ }\textbf {\bibinfo {volume}
  {84}},\ \bibinfo {pages} {023629} (\bibinfo {year} {2011})}\BibitemShut
  {NoStop}%
\bibitem [{\citenamefont {Holstein}\ and\ \citenamefont
  {Primakoff}(1940)}]{holstein1940field}%
  \BibitemOpen
  \bibfield  {author} {\bibinfo {author} {\bibfnamefont {T}~\bibnamefont
  {Holstein}}\ and\ \bibinfo {author} {\bibfnamefont {Hl}~\bibnamefont
  {Primakoff}},\ }\bibfield  {title} {\enquote {\bibinfo {title} {Field
  dependence of the intrinsic domain magnetization of a ferromagnet},}\
  }\href@noop {} {\bibfield  {journal} {\bibinfo  {journal} {Physical Review}\
  }\textbf {\bibinfo {volume} {58}},\ \bibinfo {pages} {1098} (\bibinfo {year}
  {1940})}\BibitemShut {NoStop}%
\bibitem [{\citenamefont {Ressayre}\ and\ \citenamefont
  {Tallet}(1975)}]{PhysRevA.11.981}%
  \BibitemOpen
  \bibfield  {author} {\bibinfo {author} {\bibfnamefont {E.}~\bibnamefont
  {Ressayre}}\ and\ \bibinfo {author} {\bibfnamefont {A.}~\bibnamefont
  {Tallet}},\ }\bibfield  {title} {\enquote {\bibinfo {title}
  {{Holstein-Primakoff transformation for the study of cooperative emission of
  radiation}},}\ }\href {\doibase 10.1103/PhysRevA.11.981} {\bibfield
  {journal} {\bibinfo  {journal} {Phys. Rev. A}\ }\textbf {\bibinfo {volume}
  {11}},\ \bibinfo {pages} {981--988} (\bibinfo {year} {1975})}\BibitemShut
  {NoStop}%
\bibitem [{\citenamefont {Carollo}\ and\ \citenamefont
  {Pachos}(2005)}]{carollo2005geometric}%
  \BibitemOpen
  \bibfield  {author} {\bibinfo {author} {\bibfnamefont {Angelo~CM}\
  \bibnamefont {Carollo}}\ and\ \bibinfo {author} {\bibfnamefont {Jiannis~K}\
  \bibnamefont {Pachos}},\ }\bibfield  {title} {\enquote {\bibinfo {title}
  {Geometric phases and criticality in spin-chain systems},}\ }\href@noop {}
  {\bibfield  {journal} {\bibinfo  {journal} {Phys. Rev. Lett.}\ }\textbf
  {\bibinfo {volume} {95}},\ \bibinfo {pages} {157203} (\bibinfo {year}
  {2005})}\BibitemShut {NoStop}%
\bibitem [{\citenamefont {Zanardi}\ and\ \citenamefont
  {Paunkovi\ifmmode~\acute{c}\else \'{c}\fi{}}(2006)}]{PhysRevE.74.031123}%
  \BibitemOpen
  \bibfield  {author} {\bibinfo {author} {\bibfnamefont {Paolo}\ \bibnamefont
  {Zanardi}}\ and\ \bibinfo {author} {\bibfnamefont {Nikola}\ \bibnamefont
  {Paunkovi\ifmmode~\acute{c}\else \'{c}\fi{}}},\ }\bibfield  {title} {\enquote
  {\bibinfo {title} {Ground state overlap and quantum phase transitions},}\
  }\href {\doibase 10.1103/PhysRevE.74.031123} {\bibfield  {journal} {\bibinfo
  {journal} {Phys. Rev. E}\ }\textbf {\bibinfo {volume} {74}},\ \bibinfo
  {pages} {031123} (\bibinfo {year} {2006})}\BibitemShut {NoStop}%
\bibitem [{\citenamefont {Plastina}\ \emph {et~al.}(2006)\citenamefont
  {Plastina}, \citenamefont {Liberti},\ and\ \citenamefont
  {Carollo}}]{plastina2006scaling}%
  \BibitemOpen
  \bibfield  {author} {\bibinfo {author} {\bibfnamefont {Francesco}\
  \bibnamefont {Plastina}}, \bibinfo {author} {\bibfnamefont {Giuseppe}\
  \bibnamefont {Liberti}}, \ and\ \bibinfo {author} {\bibfnamefont {Angelo}\
  \bibnamefont {Carollo}},\ }\bibfield  {title} {\enquote {\bibinfo {title}
  {{Scaling of Berry's phase close to the Dicke quantum phase transition}},}\
  }\href@noop {} {\bibfield  {journal} {\bibinfo  {journal} {EPL (Europhysics
  Letters)}\ }\textbf {\bibinfo {volume} {76}},\ \bibinfo {pages} {182}
  (\bibinfo {year} {2006})}\BibitemShut {NoStop}%
\bibitem [{\citenamefont {Giovannetti}\ \emph {et~al.}(2003)\citenamefont
  {Giovannetti}, \citenamefont {Mancini}, \citenamefont {Vitali},\ and\
  \citenamefont {Tombesi}}]{giovannetti2003characterizing}%
  \BibitemOpen
  \bibfield  {author} {\bibinfo {author} {\bibfnamefont {Vittorio}\
  \bibnamefont {Giovannetti}}, \bibinfo {author} {\bibfnamefont {Stefano}\
  \bibnamefont {Mancini}}, \bibinfo {author} {\bibfnamefont {David}\
  \bibnamefont {Vitali}}, \ and\ \bibinfo {author} {\bibfnamefont {Paolo}\
  \bibnamefont {Tombesi}},\ }\bibfield  {title} {\enquote {\bibinfo {title}
  {Characterizing the entanglement of bipartite quantum systems},}\ }\href@noop
  {} {\bibfield  {journal} {\bibinfo  {journal} {Phys. Rev. A}\ }\textbf
  {\bibinfo {volume} {67}},\ \bibinfo {pages} {022320} (\bibinfo {year}
  {2003})}\BibitemShut {NoStop}%
\bibitem [{\citenamefont {S{\o}rensen}\ \emph {et~al.}(2001)\citenamefont
  {S{\o}rensen}, \citenamefont {Duan}, \citenamefont {Cirac},\ and\
  \citenamefont {Zoller}}]{sorensen2001Nature}%
  \BibitemOpen
  \bibfield  {author} {\bibinfo {author} {\bibfnamefont {A}~\bibnamefont
  {S{\o}rensen}}, \bibinfo {author} {\bibfnamefont {L-M}\ \bibnamefont {Duan}},
  \bibinfo {author} {\bibfnamefont {JI}~\bibnamefont {Cirac}}, \ and\ \bibinfo
  {author} {\bibfnamefont {Peter}\ \bibnamefont {Zoller}},\ }\bibfield  {title}
  {\enquote {\bibinfo {title} {{Many-particle entanglement with Bose-Einstein
  condensates}},}\ }\href@noop {} {\bibfield  {journal} {\bibinfo  {journal}
  {Nature}\ }\textbf {\bibinfo {volume} {409}},\ \bibinfo {pages} {63--66}
  (\bibinfo {year} {2001})}\BibitemShut {NoStop}%
\bibitem [{\citenamefont {Baumann}\ \emph {et~al.}(2010)\citenamefont
  {Baumann}, \citenamefont {Guerlin}, \citenamefont {Brennecke},\ and\
  \citenamefont {Esslinger}}]{EsslingerNature2010Dicke}%
  \BibitemOpen
  \bibfield  {author} {\bibinfo {author} {\bibfnamefont {Kristian}\
  \bibnamefont {Baumann}}, \bibinfo {author} {\bibfnamefont {Christine}\
  \bibnamefont {Guerlin}}, \bibinfo {author} {\bibfnamefont {Ferdinand}\
  \bibnamefont {Brennecke}}, \ and\ \bibinfo {author} {\bibfnamefont {Tilman}\
  \bibnamefont {Esslinger}},\ }\bibfield  {title} {\enquote {\bibinfo {title}
  {Dicke quantum phase transition with a superfluid gas in an optical
  cavity},}\ }\href@noop {} {\bibfield  {journal} {\bibinfo  {journal}
  {Nature}\ }\textbf {\bibinfo {volume} {464}},\ \bibinfo {pages} {1301--1306}
  (\bibinfo {year} {2010})}\BibitemShut {NoStop}%
\bibitem [{\citenamefont {Riedel}\ \emph {et~al.}(2010)\citenamefont {Riedel},
  \citenamefont {B{\"o}hi}, \citenamefont {Li}, \citenamefont {H{\"a}nsch},
  \citenamefont {Sinatra},\ and\ \citenamefont {Treutlein}}]{riedel2010atom}%
  \BibitemOpen
  \bibfield  {author} {\bibinfo {author} {\bibfnamefont {Max~F}\ \bibnamefont
  {Riedel}}, \bibinfo {author} {\bibfnamefont {Pascal}\ \bibnamefont
  {B{\"o}hi}}, \bibinfo {author} {\bibfnamefont {Yun}\ \bibnamefont {Li}},
  \bibinfo {author} {\bibfnamefont {Theodor~W}\ \bibnamefont {H{\"a}nsch}},
  \bibinfo {author} {\bibfnamefont {Alice}\ \bibnamefont {Sinatra}}, \ and\
  \bibinfo {author} {\bibfnamefont {Philipp}\ \bibnamefont {Treutlein}},\
  }\bibfield  {title} {\enquote {\bibinfo {title} {Atom-chip-based generation
  of entanglement for quantum metrology},}\ }\href@noop {} {\bibfield
  {journal} {\bibinfo  {journal} {Nature}\ }\textbf {\bibinfo {volume} {464}},\
  \bibinfo {pages} {1170--1173} (\bibinfo {year} {2010})}\BibitemShut {NoStop}%
\bibitem [{\citenamefont {Matsukevich}\ \emph {et~al.}(2006)\citenamefont
  {Matsukevich}, \citenamefont {Chaneliere}, \citenamefont {Jenkins},
  \citenamefont {Lan}, \citenamefont {Kennedy},\ and\ \citenamefont
  {Kuzmich}}]{matsukevich2006entanglement}%
  \BibitemOpen
  \bibfield  {author} {\bibinfo {author} {\bibfnamefont {DN}~\bibnamefont
  {Matsukevich}}, \bibinfo {author} {\bibfnamefont {Thierry}\ \bibnamefont
  {Chaneliere}}, \bibinfo {author} {\bibfnamefont {SD}~\bibnamefont {Jenkins}},
  \bibinfo {author} {\bibfnamefont {S-Y}\ \bibnamefont {Lan}}, \bibinfo
  {author} {\bibfnamefont {TAB}\ \bibnamefont {Kennedy}}, \ and\ \bibinfo
  {author} {\bibfnamefont {Alex}\ \bibnamefont {Kuzmich}},\ }\bibfield  {title}
  {\enquote {\bibinfo {title} {Entanglement of remote atomic qubits},}\
  }\href@noop {} {\bibfield  {journal} {\bibinfo  {journal} {Physical review
  letters}\ }\textbf {\bibinfo {volume} {96}},\ \bibinfo {pages} {030405}
  (\bibinfo {year} {2006})}\BibitemShut {NoStop}%
\bibitem [{\citenamefont {Simon}\ \emph {et~al.}(2007)\citenamefont {Simon},
  \citenamefont {Tanji}, \citenamefont {Ghosh},\ and\ \citenamefont
  {Vuleti{\'c}}}]{simon2007single}%
  \BibitemOpen
  \bibfield  {author} {\bibinfo {author} {\bibfnamefont {Jonathan}\
  \bibnamefont {Simon}}, \bibinfo {author} {\bibfnamefont {Haruka}\
  \bibnamefont {Tanji}}, \bibinfo {author} {\bibfnamefont {Saikat}\
  \bibnamefont {Ghosh}}, \ and\ \bibinfo {author} {\bibfnamefont {Vladan}\
  \bibnamefont {Vuleti{\'c}}},\ }\bibfield  {title} {\enquote {\bibinfo {title}
  {Single-photon bus connecting spin-wave quantum memories},}\ }\href@noop {}
  {\bibfield  {journal} {\bibinfo  {journal} {Nature Physics}\ }\textbf
  {\bibinfo {volume} {3}},\ \bibinfo {pages} {765} (\bibinfo {year}
  {2007})}\BibitemShut {NoStop}%
\bibitem [{\citenamefont {Fadel}\ \emph {et~al.}(2018)\citenamefont {Fadel},
  \citenamefont {Zibold}, \citenamefont {D{\'e}camps},\ and\ \citenamefont
  {Treutlein}}]{fadel2018spatial}%
  \BibitemOpen
  \bibfield  {author} {\bibinfo {author} {\bibfnamefont {Matteo}\ \bibnamefont
  {Fadel}}, \bibinfo {author} {\bibfnamefont {Tilman}\ \bibnamefont {Zibold}},
  \bibinfo {author} {\bibfnamefont {Boris}\ \bibnamefont {D{\'e}camps}}, \ and\
  \bibinfo {author} {\bibfnamefont {Philipp}\ \bibnamefont {Treutlein}},\
  }\bibfield  {title} {\enquote {\bibinfo {title} {{Spatial entanglement
  patterns and Einstein-Podolsky-Rosen steering in Bose-Einstein
  condensates}},}\ }\href@noop {} {\bibfield  {journal} {\bibinfo  {journal}
  {Science}\ }\textbf {\bibinfo {volume} {360}},\ \bibinfo {pages} {409--413}
  (\bibinfo {year} {2018})}\BibitemShut {NoStop}%
\bibitem [{\citenamefont {Kunkel}\ \emph {et~al.}(2018)\citenamefont {Kunkel},
  \citenamefont {Pr{\"u}fer}, \citenamefont {Strobel}, \citenamefont
  {Linnemann}, \citenamefont {Fr{\"o}lian}, \citenamefont {Gasenzer},
  \citenamefont {G{\"a}rttner},\ and\ \citenamefont
  {Oberthaler}}]{kunkel2018spatially}%
  \BibitemOpen
  \bibfield  {author} {\bibinfo {author} {\bibfnamefont {Philipp}\ \bibnamefont
  {Kunkel}}, \bibinfo {author} {\bibfnamefont {Maximilian}\ \bibnamefont
  {Pr{\"u}fer}}, \bibinfo {author} {\bibfnamefont {Helmut}\ \bibnamefont
  {Strobel}}, \bibinfo {author} {\bibfnamefont {Daniel}\ \bibnamefont
  {Linnemann}}, \bibinfo {author} {\bibfnamefont {Anika}\ \bibnamefont
  {Fr{\"o}lian}}, \bibinfo {author} {\bibfnamefont {Thomas}\ \bibnamefont
  {Gasenzer}}, \bibinfo {author} {\bibfnamefont {Martin}\ \bibnamefont
  {G{\"a}rttner}}, \ and\ \bibinfo {author} {\bibfnamefont {Markus~K}\
  \bibnamefont {Oberthaler}},\ }\bibfield  {title} {\enquote {\bibinfo {title}
  {{Spatially distributed multipartite entanglement enables EPR steering of
  atomic clouds}},}\ }\href@noop {} {\bibfield  {journal} {\bibinfo  {journal}
  {Science}\ }\textbf {\bibinfo {volume} {360}},\ \bibinfo {pages} {413--416}
  (\bibinfo {year} {2018})}\BibitemShut {NoStop}%
\bibitem [{\citenamefont {Lange}\ \emph {et~al.}(2018)\citenamefont {Lange},
  \citenamefont {Peise}, \citenamefont {L{\"u}cke}, \citenamefont {Kruse},
  \citenamefont {Vitagliano}, \citenamefont {Apellaniz}, \citenamefont
  {Kleinmann}, \citenamefont {T{\'o}th},\ and\ \citenamefont
  {Klempt}}]{lange2018entanglement}%
  \BibitemOpen
  \bibfield  {author} {\bibinfo {author} {\bibfnamefont {Karsten}\ \bibnamefont
  {Lange}}, \bibinfo {author} {\bibfnamefont {Jan}\ \bibnamefont {Peise}},
  \bibinfo {author} {\bibfnamefont {Bernd}\ \bibnamefont {L{\"u}cke}}, \bibinfo
  {author} {\bibfnamefont {Ilka}\ \bibnamefont {Kruse}}, \bibinfo {author}
  {\bibfnamefont {Giuseppe}\ \bibnamefont {Vitagliano}}, \bibinfo {author}
  {\bibfnamefont {Iagoba}\ \bibnamefont {Apellaniz}}, \bibinfo {author}
  {\bibfnamefont {Matthias}\ \bibnamefont {Kleinmann}}, \bibinfo {author}
  {\bibfnamefont {G{\'e}za}\ \bibnamefont {T{\'o}th}}, \ and\ \bibinfo {author}
  {\bibfnamefont {Carsten}\ \bibnamefont {Klempt}},\ }\bibfield  {title}
  {\enquote {\bibinfo {title} {Entanglement between two spatially separated
  atomic modes},}\ }\href@noop {} {\bibfield  {journal} {\bibinfo  {journal}
  {Science}\ }\textbf {\bibinfo {volume} {360}},\ \bibinfo {pages} {416--418}
  (\bibinfo {year} {2018})}\BibitemShut {NoStop}%
\bibitem [{\citenamefont {Fukuhara}\ \emph {et~al.}(2015)\citenamefont
  {Fukuhara}, \citenamefont {Hild}, \citenamefont {Zeiher}, \citenamefont
  {Schau\ss{}}, \citenamefont {Bloch}, \citenamefont {Endres},\ and\
  \citenamefont {Gross}}]{PhysRevLett.115.035302}%
  \BibitemOpen
  \bibfield  {author} {\bibinfo {author} {\bibfnamefont {Takeshi}\ \bibnamefont
  {Fukuhara}}, \bibinfo {author} {\bibfnamefont {Sebastian}\ \bibnamefont
  {Hild}}, \bibinfo {author} {\bibfnamefont {Johannes}\ \bibnamefont {Zeiher}},
  \bibinfo {author} {\bibfnamefont {Peter}\ \bibnamefont {Schau\ss{}}},
  \bibinfo {author} {\bibfnamefont {Immanuel}\ \bibnamefont {Bloch}}, \bibinfo
  {author} {\bibfnamefont {Manuel}\ \bibnamefont {Endres}}, \ and\ \bibinfo
  {author} {\bibfnamefont {Christian}\ \bibnamefont {Gross}},\ }\bibfield
  {title} {\enquote {\bibinfo {title} {{Spatially resolved detection of a
  spin-entanglement wave in a Bose-Hubbard Chain}},}\ }\href {\doibase
  10.1103/PhysRevLett.115.035302} {\bibfield  {journal} {\bibinfo  {journal}
  {Phys. Rev. Lett.}\ }\textbf {\bibinfo {volume} {115}},\ \bibinfo {pages}
  {035302} (\bibinfo {year} {2015})}\BibitemShut {NoStop}%
\bibitem [{\citenamefont {Islam}\ \emph {et~al.}(2015)\citenamefont {Islam},
  \citenamefont {Ma}, \citenamefont {Preiss}, \citenamefont {Tai},
  \citenamefont {Lukin}, \citenamefont {Rispoli},\ and\ \citenamefont
  {Greiner}}]{islam2015measuring}%
  \BibitemOpen
  \bibfield  {author} {\bibinfo {author} {\bibfnamefont {Rajibul}\ \bibnamefont
  {Islam}}, \bibinfo {author} {\bibfnamefont {Ruichao}\ \bibnamefont {Ma}},
  \bibinfo {author} {\bibfnamefont {Philipp~M}\ \bibnamefont {Preiss}},
  \bibinfo {author} {\bibfnamefont {M~Eric}\ \bibnamefont {Tai}}, \bibinfo
  {author} {\bibfnamefont {Alexander}\ \bibnamefont {Lukin}}, \bibinfo {author}
  {\bibfnamefont {Matthew}\ \bibnamefont {Rispoli}}, \ and\ \bibinfo {author}
  {\bibfnamefont {Markus}\ \bibnamefont {Greiner}},\ }\bibfield  {title}
  {\enquote {\bibinfo {title} {Measuring entanglement entropy in a quantum
  many-body system},}\ }\href@noop {} {\bibfield  {journal} {\bibinfo
  {journal} {Nature}\ }\textbf {\bibinfo {volume} {528}},\ \bibinfo {pages}
  {77} (\bibinfo {year} {2015})}\BibitemShut {NoStop}%
\bibitem [{\citenamefont {Vidal}\ \emph {et~al.}(2007)\citenamefont {Vidal},
  \citenamefont {Dusuel},\ and\ \citenamefont
  {Barthel}}]{vidal2007entanglement}%
  \BibitemOpen
  \bibfield  {author} {\bibinfo {author} {\bibfnamefont {Julien}\ \bibnamefont
  {Vidal}}, \bibinfo {author} {\bibfnamefont {S{\'e}bastien}\ \bibnamefont
  {Dusuel}}, \ and\ \bibinfo {author} {\bibfnamefont {Thomas}\ \bibnamefont
  {Barthel}},\ }\bibfield  {title} {\enquote {\bibinfo {title} {Entanglement
  entropy in collective models},}\ }\href@noop {} {\bibfield  {journal}
  {\bibinfo  {journal} {Journal of Statistical Mechanics: Theory and
  Experiment}\ }\textbf {\bibinfo {volume} {2007}},\ \bibinfo {pages} {P01015}
  (\bibinfo {year} {2007})}\BibitemShut {NoStop}%
\bibitem [{\citenamefont {He}\ \emph {et~al.}(2012)\citenamefont {He},
  \citenamefont {Drummond}, \citenamefont {Olsen},\ and\ \citenamefont
  {Reid}}]{PhysRevA.86.023626}%
  \BibitemOpen
  \bibfield  {author} {\bibinfo {author} {\bibfnamefont {Q.~Y.}\ \bibnamefont
  {He}}, \bibinfo {author} {\bibfnamefont {P.~D.}\ \bibnamefont {Drummond}},
  \bibinfo {author} {\bibfnamefont {M.~K.}\ \bibnamefont {Olsen}}, \ and\
  \bibinfo {author} {\bibfnamefont {M.~D.}\ \bibnamefont {Reid}},\ }\bibfield
  {title} {\enquote {\bibinfo {title} {{Einstein-Podolsky-Rosen entanglement
  and steering in two-well Bose-Einstein-condensate ground states}},}\ }\href
  {\doibase 10.1103/PhysRevA.86.023626} {\bibfield  {journal} {\bibinfo
  {journal} {Phys. Rev. A}\ }\textbf {\bibinfo {volume} {86}},\ \bibinfo
  {pages} {023626} (\bibinfo {year} {2012})}\BibitemShut {NoStop}%
\bibitem [{\citenamefont {Hillery}\ and\ \citenamefont
  {Zubairy}(2006)}]{HZPRL2006}%
  \BibitemOpen
  \bibfield  {author} {\bibinfo {author} {\bibfnamefont {Mark}\ \bibnamefont
  {Hillery}}\ and\ \bibinfo {author} {\bibfnamefont {M.}~\bibnamefont
  {Zubairy}},\ }\bibfield  {title} {\enquote {\bibinfo {title} {Entanglement
  conditions for two-mode states},}\ }\href {\doibase
  10.1103/PhysRevLett.96.050503} {\bibfield  {journal} {\bibinfo  {journal}
  {Phys. Rev. Lett.}\ }\textbf {\bibinfo {volume} {96}},\ \bibinfo {pages}
  {050503} (\bibinfo {year} {2006})}\BibitemShut {NoStop}%
\bibitem [{\citenamefont {Milburn}\ \emph {et~al.}(1997)\citenamefont
  {Milburn}, \citenamefont {Corney}, \citenamefont {Wright},\ and\
  \citenamefont {Walls}}]{PhysRevA.55.4318}%
  \BibitemOpen
  \bibfield  {author} {\bibinfo {author} {\bibfnamefont {G.~J.}\ \bibnamefont
  {Milburn}}, \bibinfo {author} {\bibfnamefont {J.}~\bibnamefont {Corney}},
  \bibinfo {author} {\bibfnamefont {E.~M.}\ \bibnamefont {Wright}}, \ and\
  \bibinfo {author} {\bibfnamefont {D.~F.}\ \bibnamefont {Walls}},\ }\bibfield
  {title} {\enquote {\bibinfo {title} {{Quantum dynamics of an atomic
  Bose-Einstein condensate in a double-well potential}},}\ }\href {\doibase
  10.1103/PhysRevA.55.4318} {\bibfield  {journal} {\bibinfo  {journal} {Phys.
  Rev. A}\ }\textbf {\bibinfo {volume} {55}},\ \bibinfo {pages} {4318--4324}
  (\bibinfo {year} {1997})}\BibitemShut {NoStop}%
\bibitem [{\citenamefont {J\"a\"askel\"ainen}\ and\ \citenamefont
  {Meystre}(2006)}]{PhysRevA.73.013602}%
  \BibitemOpen
  \bibfield  {author} {\bibinfo {author} {\bibfnamefont {M.}~\bibnamefont
  {J\"a\"askel\"ainen}}\ and\ \bibinfo {author} {\bibfnamefont
  {P.}~\bibnamefont {Meystre}},\ }\bibfield  {title} {\enquote {\bibinfo
  {title} {Coherence dynamics of two-mode condensates in asymmetric
  potentials},}\ }\href {\doibase 10.1103/PhysRevA.73.013602} {\bibfield
  {journal} {\bibinfo  {journal} {Phys. Rev. A}\ }\textbf {\bibinfo {volume}
  {73}},\ \bibinfo {pages} {013602} (\bibinfo {year} {2006})}\BibitemShut
  {NoStop}%
\bibitem [{\citenamefont {Sakurai}\ \emph {et~al.}(2014)\citenamefont
  {Sakurai}, \citenamefont {Napolitano} \emph {et~al.}}]{sakurai2014modern}%
  \BibitemOpen
  \bibfield  {author} {\bibinfo {author} {\bibfnamefont {Jun~John}\
  \bibnamefont {Sakurai}}, \bibinfo {author} {\bibfnamefont {Jim}\ \bibnamefont
  {Napolitano}},  \emph {et~al.},\ }\href@noop {} {\emph {\bibinfo {title}
  {Modern quantum mechanics}}},\ Vol.\ \bibinfo {volume} {185}\ (\bibinfo
  {publisher} {Pearson Harlow},\ \bibinfo {year} {2014})\BibitemShut {NoStop}%
\bibitem [{\citenamefont {Zheng}(2011)}]{ZhengPRA2011ensemble_ensemble}%
  \BibitemOpen
  \bibfield  {author} {\bibinfo {author} {\bibfnamefont {Shi-Biao}\
  \bibnamefont {Zheng}},\ }\bibfield  {title} {\enquote {\bibinfo {title}
  {Dicke-like quantum phase transition and vacuum entanglement with two coupled
  atomic ensembles},}\ }\href@noop {} {\bibfield  {journal} {\bibinfo
  {journal} {Phys. Rev. A}\ }\textbf {\bibinfo {volume} {84}},\ \bibinfo
  {pages} {033817} (\bibinfo {year} {2011})}\BibitemShut {NoStop}%
\bibitem [{\citenamefont {G\"unay}\ \emph {et~al.}(2019)\citenamefont
  {G\"unay}, \citenamefont {M\"ustecapl\ifmmode \imath \else \i
  \fi{}o\ifmmode~\breve{g}\else \u{g}\fi{}lu},\ and\ \citenamefont
  {Tasgin}}]{gunay2019entanglement}%
  \BibitemOpen
  \bibfield  {author} {\bibinfo {author} {\bibfnamefont {Mehmet}\ \bibnamefont
  {G\"unay}}, \bibinfo {author} {\bibfnamefont {\"Ozg\"ur~Esat}\ \bibnamefont
  {M\"ustecapl\ifmmode \imath \else \i \fi{}o\ifmmode~\breve{g}\else
  \u{g}\fi{}lu}}, \ and\ \bibinfo {author} {\bibfnamefont {Mehmet~Emre}\
  \bibnamefont {Tasgin}},\ }\bibfield  {title} {\enquote {\bibinfo {title}
  {Entanglement criteria for two strongly interacting ensembles},}\ }\href
  {\doibase 10.1103/PhysRevA.100.063838} {\bibfield  {journal} {\bibinfo
  {journal} {Phys. Rev. A}\ }\textbf {\bibinfo {volume} {100}},\ \bibinfo
  {pages} {063838} (\bibinfo {year} {2019})}\BibitemShut {NoStop}%
\bibitem [{\citenamefont {Ho}\ and\ \citenamefont
  {Shenoy}(1996)}]{PhysRevLett.77.3276}%
  \BibitemOpen
  \bibfield  {author} {\bibinfo {author} {\bibfnamefont {Tin-Lun}\ \bibnamefont
  {Ho}}\ and\ \bibinfo {author} {\bibfnamefont {V.~B.}\ \bibnamefont
  {Shenoy}},\ }\bibfield  {title} {\enquote {\bibinfo {title} {{Binary mixtures
  of Bose condensates of alkali atoms}},}\ }\href {\doibase
  10.1103/PhysRevLett.77.3276} {\bibfield  {journal} {\bibinfo  {journal}
  {Phys. Rev. Lett.}\ }\textbf {\bibinfo {volume} {77}},\ \bibinfo {pages}
  {3276--3279} (\bibinfo {year} {1996})}\BibitemShut {NoStop}%
\bibitem [{\citenamefont {\"Ohberg}(1999)}]{PhysRevA.59.634}%
  \BibitemOpen
  \bibfield  {author} {\bibinfo {author} {\bibfnamefont {Patrik}\ \bibnamefont
  {\"Ohberg}},\ }\bibfield  {title} {\enquote {\bibinfo {title} {{Stability
  properties of the two-component Bose-Einstein condensate}},}\ }\href
  {\doibase 10.1103/PhysRevA.59.634} {\bibfield  {journal} {\bibinfo  {journal}
  {Phys. Rev. A}\ }\textbf {\bibinfo {volume} {59}},\ \bibinfo {pages}
  {634--638} (\bibinfo {year} {1999})}\BibitemShut {NoStop}%
\bibitem [{\citenamefont {Pattinson}\ \emph {et~al.}(2013)\citenamefont
  {Pattinson}, \citenamefont {Billam}, \citenamefont {Gardiner}, \citenamefont
  {McCarron}, \citenamefont {Cho}, \citenamefont {Cornish}, \citenamefont
  {Parker},\ and\ \citenamefont {Proukakis}}]{PhysRevA.87.013625}%
  \BibitemOpen
  \bibfield  {author} {\bibinfo {author} {\bibfnamefont {R.~W.}\ \bibnamefont
  {Pattinson}}, \bibinfo {author} {\bibfnamefont {T.~P.}\ \bibnamefont
  {Billam}}, \bibinfo {author} {\bibfnamefont {S.~A.}\ \bibnamefont
  {Gardiner}}, \bibinfo {author} {\bibfnamefont {D.~J.}\ \bibnamefont
  {McCarron}}, \bibinfo {author} {\bibfnamefont {H.~W.}\ \bibnamefont {Cho}},
  \bibinfo {author} {\bibfnamefont {S.~L.}\ \bibnamefont {Cornish}}, \bibinfo
  {author} {\bibfnamefont {N.~G.}\ \bibnamefont {Parker}}, \ and\ \bibinfo
  {author} {\bibfnamefont {N.~P.}\ \bibnamefont {Proukakis}},\ }\bibfield
  {title} {\enquote {\bibinfo {title} {{Equilibrium solutions for immiscible
  two-species Bose-Einstein condensates in perturbed harmonic traps}},}\ }\href
  {\doibase 10.1103/PhysRevA.87.013625} {\bibfield  {journal} {\bibinfo
  {journal} {Phys. Rev. A}\ }\textbf {\bibinfo {volume} {87}},\ \bibinfo
  {pages} {013625} (\bibinfo {year} {2013})}\BibitemShut {NoStop}%
\bibitem [{\citenamefont {Wang}\ and\ \citenamefont
  {Hioe}(1973)}]{wang1973phase}%
  \BibitemOpen
  \bibfield  {author} {\bibinfo {author} {\bibfnamefont {Yo~K}\ \bibnamefont
  {Wang}}\ and\ \bibinfo {author} {\bibfnamefont {FT}~\bibnamefont {Hioe}},\
  }\bibfield  {title} {\enquote {\bibinfo {title} {{Phase transition in the
  Dicke model of superradiance}},}\ }\href@noop {} {\bibfield  {journal}
  {\bibinfo  {journal} {Phys. Rev. A}\ }\textbf {\bibinfo {volume} {7}},\
  \bibinfo {pages} {831} (\bibinfo {year} {1973})}\BibitemShut {NoStop}%
\bibitem [{\citenamefont {Hepp}\ and\ \citenamefont
  {Lieb}(1973)}]{hepp1973superradiant}%
  \BibitemOpen
  \bibfield  {author} {\bibinfo {author} {\bibfnamefont {Klaus}\ \bibnamefont
  {Hepp}}\ and\ \bibinfo {author} {\bibfnamefont {Elliott~H}\ \bibnamefont
  {Lieb}},\ }\bibfield  {title} {\enquote {\bibinfo {title} {{On the
  superradiant phase transition for molecules in a quantized radiation field:
  the Dicke maser model}},}\ }\href@noop {} {\bibfield  {journal} {\bibinfo
  {journal} {Annals of Physics}\ }\textbf {\bibinfo {volume} {76}},\ \bibinfo
  {pages} {360--404} (\bibinfo {year} {1973})}\BibitemShut {NoStop}%
\bibitem [{\citenamefont {Emary}\ and\ \citenamefont
  {Brandes}(2003)}]{emary2003chaos}%
  \BibitemOpen
  \bibfield  {author} {\bibinfo {author} {\bibfnamefont {Clive}\ \bibnamefont
  {Emary}}\ and\ \bibinfo {author} {\bibfnamefont {Tobias}\ \bibnamefont
  {Brandes}},\ }\bibfield  {title} {\enquote {\bibinfo {title} {{Chaos and the
  quantum phase transition in the Dicke model}},}\ }\href@noop {} {\bibfield
  {journal} {\bibinfo  {journal} {Phys. Rev. E}\ }\textbf {\bibinfo {volume}
  {67}},\ \bibinfo {pages} {066203} (\bibinfo {year} {2003})}\BibitemShut
  {NoStop}%
\bibitem [{\citenamefont {Lambert}\ \emph {et~al.}(2004)\citenamefont
  {Lambert}, \citenamefont {Emary},\ and\ \citenamefont
  {Brandes}}]{lambert2004entanglement}%
  \BibitemOpen
  \bibfield  {author} {\bibinfo {author} {\bibfnamefont {Neill}\ \bibnamefont
  {Lambert}}, \bibinfo {author} {\bibfnamefont {Clive}\ \bibnamefont {Emary}},
  \ and\ \bibinfo {author} {\bibfnamefont {Tobias}\ \bibnamefont {Brandes}},\
  }\bibfield  {title} {\enquote {\bibinfo {title} {Entanglement and the phase
  transition in single-mode superradiance},}\ }\href@noop {} {\bibfield
  {journal} {\bibinfo  {journal} {Phys. Rev. Lett.}\ }\textbf {\bibinfo
  {volume} {92}},\ \bibinfo {pages} {073602} (\bibinfo {year}
  {2004})}\BibitemShut {NoStop}%
\bibitem [{\citenamefont {T{\'o}th}\ \emph {et~al.}(2009)\citenamefont
  {T{\'o}th}, \citenamefont {Knapp}, \citenamefont {G{\"u}hne},\ and\
  \citenamefont {Briegel}}]{toth2009spin}%
  \BibitemOpen
  \bibfield  {author} {\bibinfo {author} {\bibfnamefont {G{\'e}za}\
  \bibnamefont {T{\'o}th}}, \bibinfo {author} {\bibfnamefont {Christian}\
  \bibnamefont {Knapp}}, \bibinfo {author} {\bibfnamefont {Otfried}\
  \bibnamefont {G{\"u}hne}}, \ and\ \bibinfo {author} {\bibfnamefont {Hans~J}\
  \bibnamefont {Briegel}},\ }\bibfield  {title} {\enquote {\bibinfo {title}
  {Spin squeezing and entanglement},}\ }\href@noop {} {\bibfield  {journal}
  {\bibinfo  {journal} {Phys. Rev. A}\ }\textbf {\bibinfo {volume} {79}},\
  \bibinfo {pages} {042334} (\bibinfo {year} {2009})}\BibitemShut {NoStop}%
\bibitem [{\citenamefont {Julsgaard}\ \emph {et~al.}(2001)\citenamefont
  {Julsgaard}, \citenamefont {Kozhekin},\ and\ \citenamefont
  {Polzik}}]{PolzikNature2001ensemble}%
  \BibitemOpen
  \bibfield  {author} {\bibinfo {author} {\bibfnamefont {Brian}\ \bibnamefont
  {Julsgaard}}, \bibinfo {author} {\bibfnamefont {Alexander}\ \bibnamefont
  {Kozhekin}}, \ and\ \bibinfo {author} {\bibfnamefont {Eugene~S}\ \bibnamefont
  {Polzik}},\ }\bibfield  {title} {\enquote {\bibinfo {title} {Experimental
  long-lived entanglement of two macroscopic objects},}\ }\href@noop {}
  {\bibfield  {journal} {\bibinfo  {journal} {Nature}\ }\textbf {\bibinfo
  {volume} {413}},\ \bibinfo {pages} {400--403} (\bibinfo {year}
  {2001})}\BibitemShut {NoStop}%
\bibitem [{\citenamefont {Duan}(2011)}]{duan2011many_particle_entanglement}%
  \BibitemOpen
  \bibfield  {author} {\bibinfo {author} {\bibfnamefont {L-M}\ \bibnamefont
  {Duan}},\ }\bibfield  {title} {\enquote {\bibinfo {title} {{Entanglement
  detection in the vicinity of arbitrary Dicke states}},}\ }\href@noop {}
  {\bibfield  {journal} {\bibinfo  {journal} {Phys. Rev. Lett.}\ }\textbf
  {\bibinfo {volume} {107}},\ \bibinfo {pages} {180502} (\bibinfo {year}
  {2011})}\BibitemShut {NoStop}%
\bibitem [{\citenamefont {Tasgin}(2017)}]{tasgin2017many}%
  \BibitemOpen
  \bibfield  {author} {\bibinfo {author} {\bibfnamefont {Mehmet~Emre}\
  \bibnamefont {Tasgin}},\ }\bibfield  {title} {\enquote {\bibinfo {title}
  {Many-particle entanglement criterion for superradiantlike states},}\
  }\href@noop {} {\bibfield  {journal} {\bibinfo  {journal} {Phys. Rev. Lett.}\
  }\textbf {\bibinfo {volume} {119}},\ \bibinfo {pages} {033601} (\bibinfo
  {year} {2017})}\BibitemShut {NoStop}%
\bibitem [{\citenamefont {Hillery}\ \emph {et~al.}(2009)\citenamefont
  {Hillery}, \citenamefont {Dung},\ and\ \citenamefont
  {Niset}}]{PhysRevA.80.052335}%
  \BibitemOpen
  \bibfield  {author} {\bibinfo {author} {\bibfnamefont {Mark}\ \bibnamefont
  {Hillery}}, \bibinfo {author} {\bibfnamefont {Ho~Trung}\ \bibnamefont
  {Dung}}, \ and\ \bibinfo {author} {\bibfnamefont {Julien}\ \bibnamefont
  {Niset}},\ }\bibfield  {title} {\enquote {\bibinfo {title} {{Detecting
  entanglement with non-Hermitian operators}},}\ }\href {\doibase
  10.1103/PhysRevA.80.052335} {\bibfield  {journal} {\bibinfo  {journal} {Phys.
  Rev. A}\ }\textbf {\bibinfo {volume} {80}},\ \bibinfo {pages} {052335}
  (\bibinfo {year} {2009})}\BibitemShut {NoStop}%
\bibitem [{\citenamefont {Cavalcanti}\ and\ \citenamefont
  {Reid}(2006)}]{PhysRevLett.97.170405}%
  \BibitemOpen
  \bibfield  {author} {\bibinfo {author} {\bibfnamefont {E.~G.}\ \bibnamefont
  {Cavalcanti}}\ and\ \bibinfo {author} {\bibfnamefont {M.~D.}\ \bibnamefont
  {Reid}},\ }\bibfield  {title} {\enquote {\bibinfo {title} {Signatures for
  generalized macroscopic superpositions},}\ }\href {\doibase
  10.1103/PhysRevLett.97.170405} {\bibfield  {journal} {\bibinfo  {journal}
  {Phys. Rev. Lett.}\ }\textbf {\bibinfo {volume} {97}},\ \bibinfo {pages}
  {170405} (\bibinfo {year} {2006})}\BibitemShut {NoStop}%
\bibitem [{\citenamefont {Duan}\ \emph {et~al.}(2000)\citenamefont {Duan},
  \citenamefont {Giedke}, \citenamefont {Cirac},\ and\ \citenamefont
  {Zoller}}]{DGCZ_PRL2000}%
  \BibitemOpen
  \bibfield  {author} {\bibinfo {author} {\bibfnamefont {Lu-Ming}\ \bibnamefont
  {Duan}}, \bibinfo {author} {\bibfnamefont {G.}~\bibnamefont {Giedke}},
  \bibinfo {author} {\bibfnamefont {J.}~\bibnamefont {Cirac}}, \ and\ \bibinfo
  {author} {\bibfnamefont {P.}~\bibnamefont {Zoller}},\ }\bibfield  {title}
  {\enquote {\bibinfo {title} {Inseparability criterion for continuous variable
  systems},}\ }\href {\doibase 10.1103/PhysRevLett.84.2722} {\bibfield
  {journal} {\bibinfo  {journal} {Phys. Rev. Lett.}\ }\textbf {\bibinfo
  {volume} {84}},\ \bibinfo {pages} {2722--2725} (\bibinfo {year}
  {2000})}\BibitemShut {NoStop}%
\bibitem [{\citenamefont {Pirandola}\ \emph {et~al.}(2015)\citenamefont
  {Pirandola}, \citenamefont {Eisert}, \citenamefont {Weedbrook}, \citenamefont
  {Furusawa},\ and\ \citenamefont {Braunstein}}]{BraunsteinNaturePhot2015}%
  \BibitemOpen
  \bibfield  {author} {\bibinfo {author} {\bibfnamefont {Stefano}\ \bibnamefont
  {Pirandola}}, \bibinfo {author} {\bibfnamefont {Jens}\ \bibnamefont
  {Eisert}}, \bibinfo {author} {\bibfnamefont {Christian}\ \bibnamefont
  {Weedbrook}}, \bibinfo {author} {\bibfnamefont {Akira}\ \bibnamefont
  {Furusawa}}, \ and\ \bibinfo {author} {\bibfnamefont {Samuel~L}\ \bibnamefont
  {Braunstein}},\ }\bibfield  {title} {\enquote {\bibinfo {title} {Advances in
  quantum teleportation},}\ }\href@noop {} {\bibfield  {journal} {\bibinfo
  {journal} {Nature Photonics}\ }\textbf {\bibinfo {volume} {9}},\ \bibinfo
  {pages} {641--652} (\bibinfo {year} {2015})}\BibitemShut {NoStop}%
\bibitem [{\citenamefont {Degen}\ \emph {et~al.}(2017)\citenamefont {Degen},
  \citenamefont {Reinhard},\ and\ \citenamefont
  {Cappellaro}}]{degen2017quantum}%
  \BibitemOpen
  \bibfield  {author} {\bibinfo {author} {\bibfnamefont {Christian~L}\
  \bibnamefont {Degen}}, \bibinfo {author} {\bibfnamefont {F}~\bibnamefont
  {Reinhard}}, \ and\ \bibinfo {author} {\bibfnamefont {P}~\bibnamefont
  {Cappellaro}},\ }\bibfield  {title} {\enquote {\bibinfo {title} {Quantum
  sensing},}\ }\href@noop {} {\bibfield  {journal} {\bibinfo  {journal}
  {Reviews of modern physics}\ }\textbf {\bibinfo {volume} {89}},\ \bibinfo
  {pages} {035002} (\bibinfo {year} {2017})}\BibitemShut {NoStop}%
\end{thebibliography}%

\end{document}